\documentclass[aps,prd,10pt,tightenlines,amsmath,twocolumn,superscriptaddress,floatfix,eqsecnum,showpacs]{revtex4-1}

\usepackage[section]{placeins}
\usepackage{amssymb,amsmath}
\usepackage{dsfont}
\usepackage{bbm}
\usepackage[dvipsnames]{xcolor}
\usepackage{graphicx}
\usepackage{mathrsfs}
\usepackage{pictexwd}
\usepackage{dcpic}
\usepackage{amsthm}
\usepackage[pagebackref=false]{hyperref}
\usepackage[normalem]{ulem}

\providecommand{\ignore}[1]{}

\newif\ifcmnt
\cmnttrue
\ifdefined\cmntsoff\cmntfalse\fi
\ifcmnt
    \providecommand{\aucmnt}[1]{#1}

\else
    \providecommand{\aucmnt}[1]{}

\fi

\newtheorem*{definition}{Definition}
\newtheorem*{axiom}{Axiom}

\newcommand{\barK}{{\bar{K}}}
\newcommand{\supp}{\text{supp}}
\newcommand{\no}[1]{\mathord{:\mathrel{#1}:}}
\newcommand{\hQ}{\hat{Q}}
\newcommand{\scrA}{\mathscr{A}}

\newcommand{\ha}{\hat{a}}
\newcommand{\hH}{\hat{H}}
\newcommand{\hT}{\hat{T}}
\newcommand{\hP}{\hat{P}}
\newcommand{\hE}{\hat{E}}
\newcommand{\hB}{\hat{B}}
\newcommand{\hF}{\hat{F}}
\newcommand{\hf}{\hat{f}}
\newcommand{\hX}{\hat{\mathcal{X}}}
\newcommand{\hb}{\hat{b}}
\newcommand{\hn}{\hat{n}}
\newcommand{\<}[1]{\langle #1\rangle}

\newcommand{\g}{\mathrm{g}}
\newcommand{\ks}{{\vect{k}\sigma}}
\newcommand{\ksprime}{{\vect{k}'\sigma'}}
\newcommand{\dthreek}{\frac{d^3k}{(2\pi)^3}}
\newcommand{\vect}[1]{{\bf#1}}
\newcommand{\hvecE}{\hat{\vect{E}}}
\newcommand{\hvecB}{\hat{\vect{B}}}

\begin{document}

\title{Quantum Estimation of Parameters of Classical Spacetimes}

\author{T.~G. Downes}
\email{downes@physics.uq.edu.au}
\affiliation{Centre for Engineered Quantum Systems, School of Mathematics and Physics, The University of Queensland, St.~Lucia, Queensland 4072, Australia}

\author{J.~R. van Meter}
%\email{James.VanMeter@Colorado.edu}
\affiliation{Department of Mathematics, University of Colorado, Boulder, Colorado, 80309, USA}
\affiliation{National Institute of Standards and Technology, Boulder, Colorado, 80305, USA}

\author{E.~Knill}
\affiliation{National Institute of Standards and Technology, Boulder, Colorado, 80305, USA}

\author{G.~J. Milburn}
\affiliation{Centre for Engineered Quantum Systems, School of Mathematics and Physics, The University of Queensland, St.~Lucia, Queensland 4072, Australia}

\author{C.~M. Caves}
\affiliation{Centre for Engineered Quantum Systems, School of Mathematics and Physics, The University of Queensland, St.~Lucia, Queensland 4072, Australia}
\affiliation{Center for Quantum Information and Control, University of New Mexico, MSC07--4220, Albuquerque, New Mexico 87131-0001, USA}

\begin{abstract}
  We describe a quantum limit to measurement of classical spacetimes.
  Specifically, we formulate a quantum Cram{\'e}r-Rao lower bound for
  estimating the single parameter in any one-parameter family of
  spacetime metrics.  We employ the locally covariant formulation of
  quantum field theory in curved spacetime, which allows for a
  manifestly background-independent derivation. The result is an
  uncertainty relation that applies to all globally hyperbolic
  spacetimes.  Among other examples, we apply our method to detection
  of gravitational waves with the electromagnetic field as a probe,
  as in laser-interferometric gravitational-wave detectors.  Other
  applications are discussed, from terrestrial gravimetry to
  cosmology.
\end{abstract}


\maketitle

\section{Introduction}
\label{sec:intro}

The geometry of spacetime can be inferred from physical measurements
made with clocks and rulers or, more generally, with quantum fields,
sources, and detectors.  We assume that the ultimate precision achievable
is determined by quantum mechanics. In this paper we obtain
parameter-based quantum uncertainty relations that bound the precision
with which we can determine properties of spacetime in terms of
stress-energy variances.  Such uncertainty relations might
become increasingly relevant to empirical observation as, for example,
laser-interferometric gravitational-wave detectors are expected to
approach quantum-limited sensitivity across a wide bandwidth in the
near future.

An informative, high-level way to quantify the precision of
a parameter measurement is by the inverse variance
$\langle(\delta\tilde{\theta})^{2}\rangle$  of an estimator $\tilde{\theta}$.
The best precision with which we can measure a parameter is determined by the
quantum Fisher information~\cite{Helstrom,Holevo}.  For pure states, the Fisher information
reduces to a multiple of the variance $\langle(\Delta \hP)^{2}\rangle$
of an evolution operator~$\hP$  that describes how the quantum
state changes with changes in the parameter.  This determines a
parameter-based uncertainty relation~\cite{Braunstein:1994,Braunstein:1996},
\begin{equation}\label{eq:parameter-up}
  \langle(\delta
  \tilde{\theta})^{2}\rangle\langle(\Delta
  \hP)^{2}\rangle \geq \frac{\hbar^{2}}{4}\;,
\end{equation}
whose form is reminiscent of Heisenberg uncertainty relations.

Such parameter-based uncertainty relations can be applied to
parameters associated with local changes of the spacetime metric.
They are derived from a universal connection between local changes
in the metric and relative changes in the states of quantum fields
that live on the spacetime.  These changes, used to sense the spacetime
parameters, can be characterized in terms of an evolution of the
state driven by a stress-energy integral with respect to the change
in the metric, which gives the operator $\hP$.

For the universal connection between changes in states of quantum
fields and the stress-energy integrals, we rely on the locally covariant
formalism for quantum fields on curved spacetime
backgrounds~\cite{bfv}.  For this purpose, we treat gravity
classically as in general relativity, determined by a metric with
signature $(-,+,+,+)$ on a spacetime manifold.  This is treated as
a fixed background on which the quantum fields used for measurements---we
call these ``probe fields''---propagate. In particular, we do not consider
back action from the quantum fields on the metric. In any case, we
expect that such back action would transfer uncertainty in the quantum
field being measured to the metric and reduce the achievable
precision, for otherwise, by an argument given in~\cite{hannah:1977,adelman},
the uncertainty principle for matter would be violated (see also~\cite{ng:1994}
for a study of the problem of back action when measuring the structure of
spacetime).

We allow for the presence of classical fields that can propagate on
the spacetime background.  We only consider those
fields that play a direct role as sources for the quantum fields used
for measurement.  These sources are determined by classical devices
needed to implement the measurement.  For the types of measurement
considered here, the parametrized changes of the metric are
independent of these equipment-related classical fields.  In
particular, these classical fields contain no information about the
parameter of interest, and for this reason we do not model them explicitly.

Our work in this paper relies on the parameter-based uncertainty
relation~(\ref{eq:parameter-up}).  To put such uncertainty relations
in context, we consider now how the familiar Heisenberg uncertainty relations
generalize to parameter-based uncertainty relations.  The Heisenberg
uncertainty relation for position and momentum states that the product
of the uncertainties in position and momentum, for any quantum state,
must be greater than a constant.  In terms of the variances of position
and momentum, the Heisenberg uncertainty relation is written as
\begin{align}
\langle(\Delta \hat{x})^2\rangle\langle(\Delta
\hat{p})^2\rangle\geq\frac{\hbar^2}{4}\;,
  \label{eq:heisenberg}
\end{align}
where $\hbar$ is the reduced Planck's constant. The
Heisenberg uncertainty relation is derived in standard quantum
mechanics, where position and momentum are both represented as
Hermitian operators.

A similar relation, albeit with a different interpretation, exists
between time and energy:
\begin{align} \langle(\delta t)^2\rangle\langle(\Delta
\hat{H})^2\rangle\geq\frac{\hbar^2}{4}\;.
  \label{teur}
\end{align} Unlike position, time in standard quantum mechanics is a
classical parameter. The time-energy uncertainty relation is
an example of a quantum limit on parameter estimation.
One tries to estimate a classical parameter, in this case time,
by making measurements on a quantum system, a ``clock,'' whose
evolution depends on time.  In the time-energy uncertainty
relation~(\ref{teur}), $\langle(\delta t)^2\rangle$ is the classical
variance of the estimate of $t$; this classical variance arises
ultimately from quantum uncertainties in clock variables conjugate to
the Hamiltonian~$H$.

The most common way to make quantum mechanics compatible with
classical relativity is to demote position to a parameter, just like
time in the previous example.  Physical systems can then be thought of
as living on, and interacting with, the classical spacetime manifold.
In this relativistic context, the position-momentum uncertainty
relation naturally becomes a quantum limit on parameter-estimation~\cite{Braunstein:1994}.

This parameter-based approach is the natural, operational way
to think of uncertainty relations.  Nothing in the traditional Heisenberg
uncertainty relation~(\ref{eq:heisenberg}) refers directly to a measurement
of position or momentum.  In the parameter-based approach, one considers
measurements of any sort whose results are used to estimate changes in
a parameter; quantum mechanics then says, via the quantum Fisher information,
that the uncertainty of the estimate is limited by the uncertainty in
the operator that generates changes in the parameter, in a way that looks
like, but is more powerful by being operational, a traditional Heisenberg
uncertainty relation.

This approach was used by Braunstein, Caves, and
Milburn~\cite{Braunstein:1996} to develop optimal quantum estimation
for spacetime displacements in flat Minkowski spacetime.  In spacetime,
not only can one move a fixed proper distance or time, one can also boost
and rotate.  Quantum parameter estimation was thus also developed for
the parameters corresponding to these actions~\cite{Braunstein:1996}.
The results were developed with the quantized electromagnetic field as
the probe. These results show that estimates of a spacetime translation
can be made increasingly accurate as the uncertainty in the operator that
generates the translation, boost, or rotation is made very~large.

In this paper we are interested in limits on the precision of
estimates of parameters of the classical gravitational field.
In general relativity the gravitational field is a manifestation of
the geometry of spacetime, which is described by a metric. The metric
determines the length of the invariant (proper) interval between two
spacetime events according to~\cite{mtw}
\begin{align}
        ds^2=\g_{\mu\nu}(x)dx^{\mu}dx^{\nu}\;,
        \label{metric}
\end{align}
where $\g_{\mu\nu}(x)$ is the metric tensor, with indices
$\mu,\nu=0,1,2,3$ for time and the three spatial coordinates.  The
$dx^{\mu}$ are infinitesimal coordinate differences.  We assume the
Einstein summation convention, where repeated upper and lower indices
are summed over.

Before we describe the relevant quantum parameter estimation, we ask
the following: Can any insight be found by applying the Heisenberg
uncertainty relation, in its parameter-based form, directly to a
proper distance?  It was along these lines that Unruh~\cite{UNRUH:1986p358}
derived an uncertainty relation for a component of the metric tensor.
Once coordinates are chosen, there should only be quantum uncertainty
in the proper time and the proper distance.  As these are in turn
related to the metric via Eq.~(\ref{metric}), any uncertainty in the
proper distance is equivalent to uncertainty in the metric.  By
applying the Heisenberg uncertainty relation to a proper distance,
Unruh found a simple, yet insightful uncertainty relation for a single
component of the metric.  The Unruh uncertainty relation for the
$\g_{11}$ component of the metric (assuming particular Cartesian-like
coordinates), in terms of variances, is
\begin{align}
  \langle(\delta
  \g_{11})^2\rangle\langle(\Delta\hat{T}^{11})^2\rangle\geq\frac{\hbar^2}{V^2}\;,
	\label{umur}
\end{align}
where here, and henceforth, we use units such that $G=c=1$.  The
conjugate variable to $\g_{11}$ is the corresponding component of the
quantized stress-energy tensor, in this case the
pressure~$\hat{T}^{11}$ in the $x^1$ direction.  The key feature of
this uncertainty relation is the inverse proportionality to $V^2$, the
square of the four-volume of the measurement.

We provide a general framework for deriving such an uncertainty
relation, by formulating it as a problem in quantum estimation theory.
The metric $\g_{\mu\nu}(x)$ is defined for each point $x$ on the
manifold. If the quantum probe (measurement device) occupies
some four-volume $V$, then the probe's state depends on the metric at every point in that
region.  If we consider the metric to be an arbitrary function on the
manifold, then we need to estimate an infinite number of parameters to
define it completely.  Instead, we consider regions of spacetime that
can be described by metrics characterized by a single parameter
$\theta$.  For example, the Schwarzschild metric, which describes the
spacetime around a static nonrotating black hole, is defined by the
single parameter~$M$, the mass of the black hole.  The task is to
estimate this mass parameter by making measurements on physical systems
living on the spacetime manifold.

There has recently been some promising work in this
direction~\cite{Ahmadi:2014a,Ahmadi:2014b,Doukas}, focusing on quantum
probes consisting of scalar fields in Gaussian states.  Here we
present a general formalism for relativistic quantum metrology, using
arbitrary fields and states.  In so doing, we address several related
issues, which have, we believe, not previously received enough
attention in this context.  Ensuring that quantum observables in
different spacetimes measure ``the same" physical parameter is
nontrivial. If the spacetimes differ by a global perturbation, the
positions of measurement devices therein might correspondingly differ,
further complicating the issue; more generally, spacetime points in
two such spacetimes cannot unambiguously be identified with each
other.  Care must also be taken to ensure coordinate independence.

Fortunately, a framework exists for comparing quantum observables in
perturbed, classical spacetimes in a coordinate-independent
manner~\cite{bfv}.  This locally covariant framework, which was
developed in the context of algebraic quantum field theory, serves as
our starting point.  As the current work is geared more towards
physical experiments than most literature invoking algebraic quantum
field theory, it is worth a comment.  The aim of algebraic quantum
field theory is to put quantum field theory on rigorous mathematical
footing, while the aim of what might be termed pragmatic quantum field
theory is to make experimental predictions~\cite{sep,fleming}.
Strides toward connecting the two have been made
recently~\cite{bfv,sanders,benini:2011,benini:2015,fewster:2003}, and
this progress makes the current work \hbox{possible}.

The locally covariant framework directly connects the stress-energy to
the change in state associated with a compactly supported change in the
metric.  The connection is via the concept of relative Cauchy evolution
developed in~\cite{bfv} and leads directly to our bounds on measurement
precision.  An issue is that the bounds obtained are with respect to the
best observable supported in a region that can be much bigger than the
region containing the probes.  While this means that the bounds are
guaranteed to be optimistic in the sense that they suggest a
higher-than-achievable precision, we generally wish to make the relationship
between the measurement region, stress-energy variance, and measurement precision
tighter. For this we show that the metric change can be localized, provided
that the sensitivity of our measurement to the parameter of interest is not
affected. Because the locally covariant framework requires compactly supported
regions, the localization is necessary when the parameter is a global property of spacetime.

Another important issue---perhaps the most important issue for the interpretation
and relevance of our results---is that computations of the relevant stress-energy
variances can be difficult.  In most situations, however, the probe devices
introduce fields that have large mean values compared to a zero-mean reference
state, which is typically the vacuum state.  In these cases, the calculation
can be simplified by recognizing that mean-field-independent contributions
become negligible.

Once we have determined the general parameter-based uncertainty relations
connecting measurement precision to stress-energy variances, we apply them
to several situations of interest.  The first involves estimating constant
metric components and recovers the Unruh uncertainty relations. By considering
specific metric components, we also obtain the parametrized form of the Heisenberg
uncertainty relations.  Next we study in detail the problem of interferometric
gravitational-wave detection with light.  This requires the full power of
our approach.  Here we take advantage of localization by means of a ``bump''
function supported in the region of the measurement devices and use the
large-mean-field property to enable explicit calculation of a bound on
precision that depends on the amplitude of the light fields used.  This
recovers the well-known shot-noise limit, but goes beyond this limit in two
ways.  First, for the case of a wideband  mean field on top of vacuum fluctuations,
we obtain a general shot-noise bound that does not make an assumption of narrow
bandwidth detection of the probe field; in particular, we find a wideband
shot-noise limit in terms of a frequency-weighted integration over mean photon
numbers.  Second, we find that wideband squeezing gives the optimal,
sub-shot-noise sensitivity under the assumptions we are making.  We briefly
discuss other examples, including cosmological parameters and gravimetry.

Our work is organized as follows.  In Sec.~\ref{sec:estimationtheory}
we review quantum estimation theory including, in particular, the
quantum Cram\'er-Rao bound, which is the expression of parameter-based
uncertainty principles.  In Sec.~\ref{sec:approach} we review the
locally covariant approach to quantum field theory in perturbed,
classical spacetime developed by Brunetti, Fredenhagen, and
Verch~\cite{bfv}.  In Sec.~\ref{sec:perturbation} we show that
application of the Cram\'er-Rao bound discussed in
Sec.~\ref{sec:estimationtheory} to such a perturbed spacetime results
in a coordinate-independent uncertainty relation between a local
spacetime property and a quantum operator that depends on the probe
field.  In Sec.~\ref{sec:compact} we generalize this uncertainty
relation to global spacetime properties.  In
Sec.~\ref{sec:metricestimation} we consider application of the
formalism to estimation of metric components, proper time, and proper
distance in a certain class of perturbed spacetimes.  In
Sec.~\ref{sec:gw} we derive a quantum uncertainty bound on detection
of gravitational waves using the electromagnetic field as a probe, as
in laser-interferometric gravitational-wave detectors.  Then, in
Sec.~\ref{sec:otherapp}, we consider additional applications, and
finally we conclude in Sec.~\ref{sec:conclusion}.

\section{Quantum estimation theory}
\label{sec:estimationtheory}

\begin{figure}
   \includegraphics[width=3.1in]{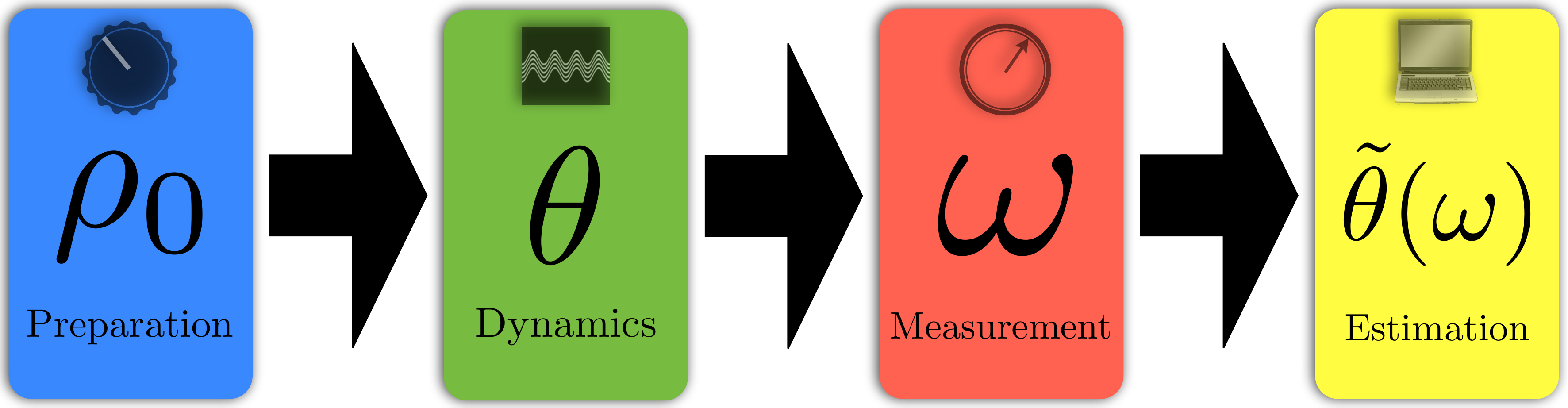}
   \caption{Scheme for quantum parameter estimation.}
   \label{qpe-scheme}
\end{figure}

In this paper we consider estimating an individual parameter of a spacetime metric, so we only need the \hbox{theory} of single-parameter quantum estimation.  A general scheme for quantum parameter estimation is depicted in Fig.~\ref{qpe-scheme}.  A quantum state, represented by a density operator $\hat\rho_0$, undergoes a unitary transformation $\hat{U}(\theta)$ that depends on the parameter~$\theta$ of interest, producing a one-parameter family of states, $\hat\rho(\theta)=\hat U(\theta)\hat\rho_0\hat U^\dagger(\theta)$.  Measurements are made on the system, with results $\omega$, which are fed into an estimator $\tilde{\theta}(\omega)$ of the parameter.

We consider generalized measurements, described by positive-operator-valued measures (POVMs). For simplicity we consider such POVMs given by a positive-operator-valued density $\hat E(\omega)$ that satisfies the completeness property
\begin{align}
\int d\omega\,\hat{E}(\omega)=\hat{1}\;,
\end{align}
where $\hat{1}$ is the identity operator. The outcomes of a particular measurement follow a probability distribution $p(\omega |\theta)$ conditional on the parameter~$\theta$. The probability distribution for the outcomes $\omega$ can be calculated as
\begin{align}
p(\omega|\theta)\,d\omega=\mathrm{Tr}\Big(\hat{E}(\omega)\hat{\rho}(\theta)d\omega\Big)\;.
\end{align}

%\cmc{A relevant example, couched in the field-theoretic language of this paper, comes from quantum optics, where the Mandel formula~\cite{Mandel} gives the POVM for photon counting as a normally ordered function of the operator
%\begin{equation}\label{eq:NTL}
%\hat{N}_{T,L}=\int dx\,R_{T,L}(x) \sum_\mu \hat{E}_\mu^{(-)}(x)\hat{E}_\mu^{(+)}(x)
%\end{equation}
%where $\hat{E}^{(\pm)}_\mu(x)$ are the positive- and negative-frequency components of the electric field mode $\mu$ and
%$R_{T,L}(x)$ describes the spatio-temporal extent of the detector.  This POVM determines the probability to count $m$ photons in the time interval $[t,t+T)$ in Minkowski spacetime.}

The problem of estimating the parameter~$\theta$ is essentially that of choosing a value $\tilde{\theta}$ to make a good estimate of $\theta$ by considering the observed $\omega$ in relation to the known probability distributions $p(\omega|\theta)d\omega$.  A common example is the maximum likelihood estimator, which is the choice of $\tilde{\theta}$ that retrospectively maximizes the probability of the observed measurement outcomes.

The variance of an unbiased estimate of the parameter~$\theta$, based on the distribution of measurement outcomes to be observed, is bounded by the classical Cram{\'e}r-Rao lower bound~\cite{Braunstein:1996},
\begin{align}\label{eq:Ftheta}
\langle(\delta \tilde{\theta})^2\rangle\geq\frac{1}{F(\theta)}\;,
\end{align}
where $F(\theta)$ is the classical Fisher information for the measurement, given by
\begin{align}
F(\theta)=\int d\omega\,\frac{1}{p(\omega|\theta)}\left(\frac{\partial p(\omega |\theta)}{\partial \theta}\right)^2\;.
\end{align}
In this paper we consider the special case where $\hat{\rho}$ is differentiable at $\theta_{0}$, with the differential generated by the self-adjoint operator $\hat{h}$, as expressed by
\begin{align}
\frac{d\hat{\rho}}{d\theta} =-\frac{i}{\hbar}[\hat{h},\hat{\rho}]\;,
\end{align}
Informally, we write
\begin{align}
\hat{\rho}(\theta_0+d\theta) = e^{-id\theta\hat{h}/\hbar}\hat{\rho}(\theta_0)e^{id\theta\hat{h}/\hbar}\;.
\label{inf_change_rho}
\end{align}
It can then be shown that for any POVM, the classical Fisher information satisfies $F(\theta)\le4\langle(\Delta\hat{h})^2/\hbar^2$~\cite{Helstrom,Holevo,Braunstein:1994}, where $\langle(\Delta\hat{h})^2\rangle$ is the quantum variance of the generator $\hat{h}$; moreover, there is a POVM that saturates this bound when $\hat{\rho}(\theta_0)$ is pure ~\cite{Braunstein:1994}.  Applying this bound to Eq.~(\ref{eq:Ftheta}) yields the quantum Cram{\'e}r-Rao lower bound~\cite{Braunstein:1996},
\begin{align}
\langle(\delta \tilde{\theta})^2\rangle\langle(\Delta\hat{h})^2\rangle\geq\frac{\hbar^2}{4}\;.
\end{align}
One can now construct examples by identifying parameters and their
corresponding generators. For example, since the Hamiltonian is the
generator of time translations, this gives the time-energy uncertainty
relation~(\ref{teur}) presented in the Introduction.  Since the
momentum operator generates displacements, using it as the generator
gives the parameter-based version of the Heisenberg uncertainty
relation.  Another important example is provided by the number
operator and phase, which is the basis of Heisenberg-limited phase
estimation~\cite{Holland:1993,Giovannetti:2006,lang:2014}.

\section{Relative Cauchy evolution}
\label{sec:approach}

We employ the locally covariant formulation of quantum field theory in
curved spacetime as developed by Brunetti, Fredenhagen, and
Verch~\cite{bfv}.  The approach has been used to develop a notion of
``identical physics'' on different
spacetimes~\cite{fewster:2012}.  A key result is a method for
calculating how quantum observables respond to local changes in the
background spacetime. Say we believe some particular region of the
universe to be well described by a metric $\g_{\mu\nu}^{(s)}$ that
depends on a parameter~$s$. If $s$ is assumed to parametrize a
compactly supported perturbation, the locally covariant approach can
be used to calculate how any observable $\hat{E}(s)$ responds to such
a change. The response is evaluated as the rate of change of the
observable with respect to the parameter. As we noted in the
Introduction, this is just what we need to calculate the quantum
Cram{\'e}r-Rao lower bound.

We emphasize, however, that only compactly supported perturbations in
spacetime are considered in~\cite{bfv}.  The motivation for this
restriction is similar to that for restricting the domain of
distributions to test functions and ensures that relevant quantities
are well defined.  We consider how to approach more general
perturbations in Sec.~\ref{sec:perturbation}.

The locally covariant approach is formulated in a category-theory
framework.  It involves the category of globally
hyperbolic spacetimes and the mapping of each to an algebra of
observables.  By this formalism, which is summarized in 
Appendix~\ref{sec:categorytheory}, the evolution of an observable in
response to a spacetime perturbation is made mathematically well
defined.

Note that here a spacetime is a pair $(M,\g)$, where $M$ is a
4-manifold admitting a Lorentzian metric and $\g$ is a Lorentzian
metric.  The additional property of global hyperbolicity is a
restriction on the causal structure on the manifold.  It removes the
possibility of closed time-like curves and ensures the spacetime can
be foliated into Cauchy surfaces.  This in turn ensures that any
hyperbolic field equation (Klein-Gordon, Maxwell, etc.) has a
well-posed initial-value formulation.

A one-parameter family of spacetimes $\{M,\g^{(s)}\}$ was considered in \cite{bfv}, all sharing ``initial" and ``final'' Cauchy surfaces, as well as respective neighborhoods $N_-$ and $N_+$ of those Cauchy surfaces.  These spacetimes differ only in their geometry, and only within a compact region between
$N_-$ and $N_+$.  We refer to the metric $\g^{(s)}$ as {\it perturbed\/} when $s\neq 0$ and {\it fiducial\/} when $s=0$.

To be more precise, we make the following geometric assumptions:
\begin{enumerate}
        \item $(M,\g^{(0)})$ is a globally hyperbolic spacetime.
        \item We choose a Cauchy surface $C$ in $(M,\g^{(0)})$ and two open subregions $(N_{\pm},\g^{(0)}_{N_\pm})$ with the following properties:
                \begin{itemize}
                        \item $N_+$ is within the future and $N_-$ is within the past causal regions of the Cauchy surface $C$.
                        \item $(N_{\pm},\g^{(0)}_{N_\pm})$ are globally hyperbolic spacetimes.
                        \item $N_{\pm}$ contain Cauchy surfaces for the whole spacetime $(M,\g^{(0)})$.
                \end{itemize}
        \item $\{\g^{(s)}\}_{s\in[-1,1]}$ is a set of Lorentzian metrics on $M$ with the following properties:
                \begin{itemize}
                        \item Each $\g^{(s)}$ deviates from $\g^{(0)}$ only on a compact subset of the region in the past of $N_+$ and the future of $N_-$.
                        \item Each $(M,\g^{(s)})$ is a globally hyperbolic spacetime.
                        \item $C$ is also a Cauchy surface for each $(M,\g^{(s)})$.
                \end{itemize}
\end{enumerate}
The geometric assumptions listed here can be seen in greater mathematical detail in Sec.~4.1 of~\cite{bfv}.

Consider now a Hilbert-space operator $\hat{A}(0)$ defined on $(M,g^{(0)})$;
this is an operator acting on a representation of the algebra generated by
the quantum fields on $(M,g^{(0)})$.  This operator could, for example, be
a POVM element for a particle detector and belong to the algebra of
operators localized to the spatiotemporal extent where the detector
is active.
% as in the photon-counting example surrounding Eq.~(\ref{eq:NTL}).}

It was shown in \cite{bfv} that $\hat{A}(0)$ unitarily transforms under an $s$-parametrized metric perturbation into a new operator $\hat{A}(s)$.  The functional derivative of the action of this unitary transformation with respect to the metric is defined as
\begin{align}
	\left.\frac{d}{ds}\right|_0\hat{A}(s) =\int_{M}d\mathring\mu(x)\,\frac{\delta\hat{A}(s)}{\delta \g_{\mu\nu}(x)}\left.\frac{d}{ds}\right|_0 \g^{(s)}_{\mu\nu}(x)\;,
	\label{BFVeq13}
\end{align}
where $d\mathring\mu(x)=\sqrt{|\det \g^{(0)}|}\,dx^0\,dx^1\,dx^2\,dx^3$ is the proper volume element for the metric $\g^{(0)}$.  The interpretation of this is as follows: Fields are prepared in $N_-$, they then scatter off an intermediate region (the compact subset of geometric assumption~3) and are measured in $N^+$; $s$ controls a localized perturbation within this region, and Eq.~(\ref{BFVeq13}) gives the infinitesimal movement of the observables in the Hilbert-space representation of $\mathscr{A}(M,\g^{(0)})$ due to an infinitesimal perturbation $ds$ around $s=0$. Further properties and interpretations of this functional derivative and the relative Cauchy evolution can be found in~\cite{bfv,fewster:2012,fewster:2011}.

For the case of the Klein-Gordon field, with its corresponding Weyl algebra of observables, it can be shown that both elements of this algebra and polynomials of field operators constructed from it obey the following relation (Theorem 4.3 from~\cite{bfv}):
\begin{align}
        \left.\frac{\delta\hat{A}(s)}{\delta \g_{\mu\nu}(x)}\right|_{s=0} = \frac{i}{2\hbar}\big[\hat{A}(0),\hat{T}^{\mu\nu}(x)\big]\;.
        \label{remarkC43}
\end{align}
Here $\hat{T}^{\mu\nu}(x)$ is the renormalized stress-energy tensor on the relevant Hilbert space as discussed in \cite{bfv} and satisfying Theorem~4.6.1 of \cite{wald}. We assume, more specifically, that it is normally ordered in accordance with the procedure advocated by Brown and Ottewill~\cite{BO}.

Inserting Eq.~(\ref{remarkC43}) into Eq.~(\ref{BFVeq13}), we have
\begin{align}
        \left.\frac{d}{ds}\right|_0\hat{A}(s) = \int_{M}d\mathring\mu(x)\,\frac{i}{2\hbar}\big[\hat{A}(0),\hat{T}^{\mu\nu}(x)\big]
        \left.\frac{d}{ds}\right|_0 \g^{(s)}_{\mu\nu}(x)\;,
\end{align}
where we emphasize that the operator $\hat{A}$ does not depend on~$x$.  By defining the operator $\hat{P}$ as
\begin{align}
        \hat{P} =\frac{1}{2}\int_{M}d\mathring\mu(x)\,\hat{T}^{\mu\nu}(x)\left.\frac{d}{ds}\right|_0\g^{(s)}_{\mu\nu}(x)
        \label{bigp}\;,
\end{align}
we have
\begin{align}
        \left.\frac{d}{ds}\hat{A}(s)\right|_{s=0} =\frac{i}{\hbar}\big[\hat{A}(0),\hat{P}\big]\;.
        \label{mainBFVresult}
\end{align}

The above relative Cauchy evolution equation has also been shown to hold for spin-$\frac{1}{2}$ and spin-1 fields \cite{sanders,benini:2011,benini:2015}, with an appropriately defined stress-energy tensor.  For example, for the electromagnetic field, which we consider in Sec.~\ref{sec:gw}, the Weyl algebra of the Klein-Gordon field is simply replaced by the Weyl algebra of gauge-equivalence classes of the vector potential.  Then by Theorem~3.2.9 in \cite{benini:2011}, the functional derivative with respect to the perturbed metric of an operator $\hat{A}$ is again given by the commutator with the stress-energy tensor, as in Eq.~(\ref{remarkC43}).

\section{Estimation of spacetime perturbation}
\label{sec:perturbation}

Due to the dual nature of operators and states, a consequence of Eq.~(\ref{mainBFVresult}) is that we can write
\begin{align}
        \hat{\rho}(0+ds) =e^{-ids\hat{P}/\hbar}\hat{\rho}(0)e^{ids\hat{P}/\hbar}\;,
        \label{sderiv}
\end{align}
where $\hat{\rho}(s)$ is a density operator in the Gelfand-Neimark-Segal~\cite{gelfand,segal,haag:1996} representation of the algebra on $(M,\g^{(0)})$ after the action of $\hat{U}(s)$, the unitary in the Hilbert-space representation corresponding to the relative Cauchy evolution (more precisely to a unit-preserving automorphism on the algebra of observables, called $\beta_{\g^{(s)}}$ in Appendix~\ref{sec:categorytheory}).  Then, noting the equivalence of Eq.~(\ref{sderiv}) to Eq.~(\ref{inf_change_rho}), we obtain~\cite{downesThesis}
\begin{align}\label{eq:qcrb}
\langle(\delta \tilde{s})^2\rangle\langle(\Delta \hat{P})^2\rangle\geq\frac{\hbar^2}{4}\;.
\end{align}
where $\tilde{s}$ is the estimator for the perturbation parameter.
%\cmc{Throughout the following, the stress-energy tensor is assumed to be normally ordered, in accordance with the procedure advocated by Brown and Ottewill~\cite{BO}.}

A limitation of the quantum Cram\'er-Rao bound~(\ref{eq:qcrb}) is that
the spacetime perturbation is restricted to compact support, whereas
physically interesting perturbations are typically not so restricted.
For example, variation in the mass of the Earth would vary the metric
at unbounded distances away.  To apply our formalism to this situation,
we approximate global perturbations compactly. As a first attempt, one
might consider using compact perturbations that approach the global one
in some limit, but this might lead to unbounded values of
$(\Delta\hat{P})^{2}$, which would trivialize the bound~(\ref{eq:qcrb}).
To avoid this we take advantage of the fact that the measurements that
yield an estimate of $\tilde\theta$ are of observables that are accessible
in a compact measurement region determined by the measurement device.
Given this, the bounds above are necessarily conservative for perturbations
with large extent: They apply to all observables in the region of the
perturbation, even those causally separated from the measurement device.
We can take advantage of a flexibility built into quantum estimation theory
whereby we can obtain an uncertainty bound from any parameter that our
estimator $\tilde\theta$ is sensitive to.

To see this, let $\Theta$ denote the global parameter of interest for a
spacetime with metric $g_{\mu\nu}(\Theta)$, with $\Theta_0$ being the
fiducial value of the parameter and $\tilde\theta$ denoting its estimator.
We can choose a smooth, compactly supported ``bump'' function,
$0\leq\chi\leq1$, with $\chi(x)=1$ on the measurement region. We consider
the localized perturbation $\g_{\mu\nu}(\theta)\equiv
g_{\mu\nu}\big(\theta_0+(\theta-\theta_0)\chi\big)$, parametrized in terms of
$\theta$, with $\theta_{0}=\Theta_{0}$.  If $\chi$ has sufficiently large
extent and transitions to $0$ sufficiently slowly (say, adiabatically),
we can argue that the sensitivity of $\tilde\theta$ to $\theta$, given by
$(d\langle\tilde\theta_{\theta}\rangle/d\theta)|_{\theta=\theta_0}$ approaches
that of $\tilde\theta$ to $\Theta$, which is
$(d\langle\tilde\theta\rangle_{\Theta}/d\Theta)|_{\Theta=\Theta_0}=1$, where
the latter identity follows from the assumption that $\tilde\theta$ is an
unbiased estimator to first order in $\Theta-\Theta_{0}$.  This means that
$\tilde\theta$ is also an unbiased estimator of $\theta$, to first order in
$\theta-\theta_{0}$, so that the Cram\'er-Rao bound applies to $\tilde\theta$
with $\g_{\mu\nu}(\theta)$, giving
\begin{equation}
    \<{(\delta\tilde{\theta})^2}\<{(\Delta\hP)^2}\geq\frac{\hbar^2}{4}\;,
    \label{thur}
\end{equation}
where
\begin{equation}
    \hP=\frac{1}{2}\int_Md\mathring{\mu}\,\hT^{\mu\nu}\left.\frac{d}{d\theta}\right|_{\theta_0}\g_{\mu\nu}(\theta)\;.
    \label{Ptheta}
\end{equation}

How the bump function transitions from $1$ on the measurement
region to $0$ is arbitrary.  The choice affects the bound, however,
through excess contributions to the variance $\<{(\Delta\hP)^2}$ in
the bump function's support outside the measurement region. To get
the best bounds on precision, we choose bump functions that minimize
this excess variance while achieving the desired sensitivity.  In
the examples to be considered, this excess variance can be
attributed to contributions from a reference state such as the
Minkowski vacuum. This is because the state associated with the
measurement and from which the bound is computed is a localized
deviation from the reference state and the bump function
necessarily extends beyond the region of localization.  We observe
that the shape of the transition of the bump function from $1$ to
$0$ affects the excess variance~\cite{borgman,satz,fewster:2015}.
In particular, the excess can be reduced by ensuring that the
transition from $1$ to $0$ is slow. This is analogous to the
adiabatic limit. For the case of the Minkowski vacuum, the reference-state
contribution can be made arbitrarily small by this method, as
demonstrated for example in~\cite{borgman}.

The use of slowly varying bump functions is expected to reduce
excess variance from the reference state, but does not necessarily
lead to readily computable bounds. For this, we observe that
informative measurements rely on deviations from the reference state
with relatively large localized mean fields.  This is both out of
necessity and to maximize the signal to noise. Typical measurements
are designed not to detect the reference-state contributions to the
variance, but rather to detect an effect in the presence of a strong
mean field, which greatly enhances the signal we are looking for. Indeed,
for arbitrary curved spacetimes, it is not known how to calculate the
reference-state contributions, nor is it known how to design a
measurement on the probe field that detects the corresponding
mean-field-independent effects.  Neglect of reference-state contributions
can then be regarded as a way of finding quantum limits on the kinds
of measurements we know how to do, which involve large mean fields.

For the case
of free fields, the large-mean-field scenario is formalized by considering the
measurement state as a displacement by a local Weyl unitary of a
reference state with mean field zero. In Minkowski space, these
displacements are enacted by conventional modal displacement
operators, where the displacement is by an amount determined by the
mean field.  We show in Sec.~\ref{sec:gw} that the variance
$\<{(\Delta\hP)^2}$ has terms that grow with the mean field as well
as mean-field independent terms.  We identify the mean-field-independent terms as
the reference-state contribution to the variance.  For a fixed bump
function, but large mean field, the reference-state contribution
becomes negligible.  This is the main strategy used for the analysis
of gravitational-wave detection in Sec.~\ref{sec:gw}.

For the above discussion, we assumed that the measurement device
is contained in a finite measurement region, where an incoming
reference state such as the vacuum state is temporarily modified for
the measurement.  This modification is usually necessary to enhance
the signal that we are looking for. In the case of an
interferometric measurement using light, the modification is
accomplished by introducing a large amplitude light field confined
between mirrors.  To accommodate these modifications in the
generally covariant formalism, the background includes externally
introduced classical sources with fixed relationships to the manifold,
meaning that these relationships are unchanged by the perturbation of
the metric under investigation. The formalism and relative
Cauchy evolution still applies, as suggested in~\cite{bfv}.
For the example of gravitational-wave detection, the fixed relationship
can be justified by the observation that the classical sources
follow geodesics for the original as well as the perturbed metric
and are thus manifestly independent of the perturbation in
Gaussian normal coordinates.

While our examples involving large mean-field deviations are on flat
backgrounds, we expect that the justification for neglecting reference-state
contributions based on large mean-field deviations also applies
to general curved backgrounds.  For any free field in a (globally
hyperbolic) spacetime, there exist zero-mean-field reference
states, called Hadamard states, characterized by
well-defined two-point correlators~\cite{wald, sanders, fewster:2003}.
Such a reference state allows us to perform normal ordering via the
point-splitting approach.  A corresponding stress-energy tensor can
then be constructed, with a well-defined expectation
value~\cite{wald}, which is unique up to terms which cancel in the
commutator~(\ref{mainBFVresult})) and in the variance (and is thus
sufficient for our purposes).  Moreover, while Hadamard reference
states are generally not unique, their contribution to the relevant
variance is mean field independent.

\section{Coordinate independence and compact perturbations}
\label{sec:compact}

The formulation given so far is generally covariant. Consider two
compactly supported perturbations of the metric where one is
obtained from the other by a local isometry, that is one acting as
the identity except on a compact region in the past of $N_{+}$ and
the future $N_{-}$.  Then both perturbations induce the same relative
Cauchy evolution, as shown in~\cite{bfv} (see also
Appendix~\ref{sec:coordindep}).  A complication to this obvious conclusion
of general covariance arises, however, when the parameter of interest
is expressed in a coordinate-dependent way and we require the use of a
bump function to localize the associated metric perturbation.  This is
the situation when estimating global parameters.  An example is the
invariant mass of a black hole, where there are a number of different
standard coordinate systems to choose from.  When the bump function
is expressed in the first coordinate system so as to be independent of
the invariant mass, in the second it can depend on the invariant mass.
This means that in the second coordinate system, the bump-function-modified
metric perturbation includes a term coming from the derivative of the
mass-dependent bump-function with respect to mass, and this leads to
discrepancies in the values of the variances and Cram\'er-Rao bounds
depending on which coordinate system is used to define the bump
function. If our sensitivity argument for the choice of bump function
is valid,  we should obtain valid bounds regardless of coordinate system.
It is desirable, however, to choose bump functions for which the dependence
on coordinate system is negligible.  In this section, we show that such
is the case for the variance of $\hP$ in an arbitrary spacetime, assuming
a sufficiently large mean field.  To demonstrate the basic
mechanism by which coordinate independence is achieved, we first consider
as an  illustrative example the expectation of $\hP$ in Schwarzschild spacetime.

\begin{widetext}
The fiducial metric in Schwarzschild coordinates is
\begin{align}
\label{schwarzschild}
\g^{S}_{\mu\nu}dx^{\mu}_Sdx^{\nu}_S
= -\biggl(1-\frac{2m_0}{r}\biggr)dt^2+\biggl(1-\frac{2m_0}{r}\biggr)^{\!-1}dr^2+r^2d\Omega^2\;,
\end{align}
and in isotropic coordinates it is
\begin{equation}
\label{isotropic}
\g^I_{\mu\nu}dx^\mu_Idx^\nu_I
= -\biggl(\frac{1-m_0/2\rho}{1+m_0/2\rho}\biggr)^2dt^2+\biggl(1+\frac{m_0}{2\rho}\biggr)^{\!4}(d\rho^2+\rho^2d\Omega^2)\;.
\end{equation}
By our prescription for approximating a global perturbation by a compact perturbation, we add to every instance of the fiducial mass $m_0$ a bump function of the form $(m-m_0)\chi(t,r,\theta,\phi)$ or $(m-m_0)\chi(t,\rho,\theta,\phi)$:
\begin{align}
\label{schwarzschildbump}
\g^{S}_{\mu\nu}dx^{\mu}_Sdx^{\nu}_S
&=-\biggl(1-\frac{2(m_0+(m-m_0)\chi)}{r}\biggr)dt^2
+\biggl(1-\frac{2(m_0+(m-m_0)\chi)}{r}\biggr)^{-1}dr^2+r^2d\Omega^2\;,\\
\label{isotropicbump}
\g^I_{\mu\nu}dx^\mu_Idx^\nu_I
&=-\left(\frac{1-[m_0+(m-m_0)\chi]/2\rho}{1+[m_0+(m-m_0)\chi/2\rho]}\right)^{\!2} dt^2
+\biggl(1+\frac{m_0+(m-m_0)\chi}{2\rho}\biggr)^{\!4}(d\rho^2+\rho^2d\Omega^2)\;.
\end{align}

The resulting metrics $\g^S_{\mu\nu}(m)$ and $\g^I_{\mu\nu}(m)$ are
not related by a coordinate transformation and thus are no longer
physically equivalent.  This reflects the fact that there is no unique
way to approximate a global perturbation with a compact perturbation.
Our particular choice depends on our initial coordinates, out of
convenience.  These metrics are, however, locally related by a
coordinate transformation on a patch restricted to the region where
$\chi=1$.  Therein, both metrics are locally indistinguishable from
that of a black hole with mass $m$ and are related by the
$m$-dependent coordinate transformation that relates Schwarzschild and
isotropic coordinates.

Now letting $m_0=0$ for simplicity, consider the (normally ordered)
expectation value of $\hP$, where the nonzero expectation value of the
stress-energy is assumed to be confined to a region~$K$ in which $\chi=1$,
i.e., $K=\supp(\<{\no{\hT^{\mu\nu}}})$ and $\chi(x)=1$ for $x\in K$.
Note that $K$ is strictly contained in the interior of $\supp(\chi)$,
since as discussed in the previous section we assume the transition of
the bump $\chi$ from 1 to 0 is both smooth and gradual.  We have
\begin{equation}
\<{\no{\hP}}=\frac{1}{2}\int_Kd\mathring\mu\,\<{\no{\hT^{\mu\nu}}}\left.\frac{d}{dm}\right|_0\g_{\mu\nu}(m)\;.
\end{equation}
Notice that $\g^S_{\mu\nu}(0)=\g^I_{\mu\nu}(0)$, since both Schwarzschild and isotropic coordinates reduce to standard spherical coordinates in this limit.  Yet
\begin{equation}
\left.\frac{d}{dm}\right|_0\g^S_{\mu\nu}(m)\neq \left.\frac{d}{dm}\right|_0\g^I_{\mu\nu}(m)\;,
\end{equation}
nor are these two tensor fields related by any coordinate transformation.
To understand this, observe that $\g_{\mu\nu}^S(m)-\g_{\mu\nu}^S(0)$
and $\g_{\mu\nu}^I(m)-\g_{\mu\nu}^I(0)$ also represent distinct tensor fields; the first terms in the two tensor fields can be made equal by an $m$-dependent coordinate transformation, but not without making the second terms unequal.  It should come as no surprise, then,
that $\<{\no{\hP_S}}$ and $\<{\no{\hP_I}}$ appear to be unequal:
\begin{align}
\langle\no{\hat{P}_S}\rangle
&=\int_K\frac{1}{r}\Bigl(\<{\no{\hat{T}^{tt}}}
+\<{\no{\hat{T}^{rr}}}\Bigr)r^2\sin\vartheta\,dt\,dr\,d\vartheta\,d\varphi\;,\\
\langle\no{\hat{P}_I}\rangle
&=\int_K\frac{1}{r}\Bigl(\<{\no{\hat{T}^{tt}}}+\<{\no{\hat{T}^{rr}}}
+r^2\<{\no{\hat{T}^{\vartheta\vartheta}}}+r^2\sin^2\!\vartheta\<{\no{\hat{T}^{\varphi\varphi}}}\Bigr)
r^2\sin\vartheta\,dt\,dr\,d\vartheta\,d\varphi\;.
\end{align}
This appearance is deceptive, however, as we see from
\begin{align}
\int_K\langle\no{\nabla_\alpha\hat{T}^{\alpha r}}\rangle\,r^2\sin\vartheta\,dt\,dr\,d\vartheta\,d\varphi
=\int_{\partial K}d\lambda\,n_\alpha\langle\no{\hat T^{\alpha r}}\rangle
-\int_K\Big(r\<{\no{\hat{T}^{\vartheta\vartheta}}}+r\sin^2\!\vartheta\<{\no{\hat{T}^{\varphi\varphi}}}\Big)
\,r^2\sin\vartheta\,dt\,dr\,d\vartheta\,d\varphi\;,
\label{eq:delTintegral}
\end{align}
where $d\lambda$ is the surface element induced on the boundary $\partial K$ of $K$ and $n_\mu$ is the corresponding surface normal.  Since $\<{\no{\hT^{\mu\nu}}}$ vanishes on $\partial K$ and assuming $\nabla_\mu\hT^{\mu\nu}=0$ (as required for a properly defined stress-energy tensor \cite{wald}), which implies that the left-hand side of Eq.~(\ref{eq:delTintegral}) vanishes identically, we conclude that
\begin{align}
\<{\no{\hP_I}}-\<{\no{\hP_S}}
=\int_K\Big(r\<{\no{\hat{T}^{\vartheta\vartheta}}}+r\sin^2\!\vartheta\<{\no{\hat{T}^{\varphi\varphi}}}\Big)
r^2\sin\vartheta\,dt\,dr\,d\vartheta\,d\varphi
=0\;.
\end{align}
Note the critical role played by the vanishing divergence of $\hT^{\mu\nu}$ in the above demonstration of coordinate independence.  This is no coincidence; for more discussion of the relevance of the stress-energy tensor to diffeomorphism invariance in the context of relative Cauchy evolution, see~\cite{bfv}.

In the above example, we only considered the first moment of $\hP$, whereas the Cram\'er-Rao bound involves the variance of $\hP$.  We deal with this question now by considering an arbitrary $s$-dependent coordinate transformation from unprimed coordinates to primed coordinates, on a coordinate patch that is assumed to cover the support $K$ of the mean stress energy (for a more general treatment see Appendix~\ref{sec:coordindep}).  In classic index notation, the metric components in the two systems are related by
\begin{align}
\g^{(s)}_{\alpha'\beta'}(x')={L^\mu}_{\alpha'}(s){L^\nu}_{\beta'}(s)\g^{(s)}_{\mu\nu}(x)\;,
\end{align}
where ${L^{\mu}}_{\alpha'}(s)=\partial x^\mu(x',s)/\partial x^{\alpha'}$.  We are interested in the tensor fields
\begin{align}
\left.\frac{\partial}{\partial s}\right|_0\g^{(s)}_{\alpha\beta}(x)
\qquad\textrm{and}\qquad
\left.\frac{\partial}{\partial s}\right|_0\g^{(s)}_{\alpha'\beta'}(x')\;,
\end{align}
where we now write the $s$-derivatives as partial derivatives to emphasize that the respective coordinates are held fixed while taking the $s$-derivative.  Because the coordinate transformation is $s$ dependent, these two tensors are not the same, but are related by
\begin{align}
\begin{split}
\left.\frac{\partial}{\partial s}\right|_0\g^{(s)}_{\alpha'\beta'}(x')
&={L^\mu}_{\alpha'}(0){L^\nu}_{\beta'}(0)
\left(\left.\frac{\partial}{\partial s}\right|_0\g^{(s)}_{\mu\nu}(x)
+\frac{\partial X^\gamma}{\partial x^\mu}\g^{(0)}_{\gamma\nu}
+\frac{\partial X^\gamma}{\partial x^\nu}\g^{(0)}_{\mu\gamma}
+X^\gamma\frac{\partial\g^{(0)}_{\mu\nu}}{\partial x^\gamma}\right)\\
&={L^\mu}_{\alpha'}(0){L^\nu}_{\beta'}(0)
\left(\left.\frac{\partial}{\partial s}\right|_0\g^{(s)}_{\mu\nu}(x)
+\nabla_\nu X_\mu+\nabla_\mu X_\nu\right)\\
&={L^\mu}_{\alpha'}(0){L^\nu}_{\beta'}(0)
\left.\frac{\partial}{\partial s}\right|_0\g^{(s)}_{\mu\nu}(x)
+\nabla_{\beta'}X_{\alpha'}+\nabla_{\alpha'}X_{\beta'}
\;,
\end{split}
\end{align}
where
\begin{align}
X^\gamma=\left.\frac{\partial}{\partial s}\right|_0 x^\gamma(x',s)\;.
\end{align}

Now we find
\begin{align}
\begin{split}
\int_K d\mathring\mu'\;\hat{T}^{\alpha'\beta'}\!\left.\frac{\partial}{\partial s}\right|_0\g^{(s)}_{\alpha'\beta'}
&=\int_K d\mathring\mu\,\hat{T}^{\mu\nu}\left(\left.\frac{\partial}{\partial s}\right|_0\g^{(s)}_{\mu\nu}+\nabla_\nu X_\mu+\nabla_\mu X_\nu\right)\\
&=\int_K d\mathring\mu \left(\hat{T}^{\mu\nu}\left.\frac{\partial}{\partial s}\right|_0\g^{(s)}_{\mu\nu}
+2\nabla_\mu(\hat{T}^{\mu\nu}X_\nu)-2(\nabla_\mu\hat{T}^{\mu\nu})X_\nu\right)\\
&=\int_K d\mathring\mu\,\hat{T}^{\mu\nu}\left.\frac{\partial}{\partial s}\right|_0\g^{(s)}_{\mu\nu}
+2\int_{\partial K} d\lambda\,n_\mu\hT^{\mu\nu}X_\nu\;,
\end{split}
\label{diffeo}
\end{align}
where in the last line we assume that $\nabla_\mu\hT^{\mu\nu}=0$ and where we convert a volume integral over $K$ to a surface integral over the boundary $\partial K$.

As before, let $K=\supp(\<{\no{\hT^{\mu\nu}}})$, so that $\<{\no{\hT^{\mu\nu}}}|_{\partial K}=0$, and we assume that $\chi|_K=1$ (and thus $K$ is strictly contained within $\supp(\chi)$).  In addition, consider a $\theta_0$-dependent coordinate transformation of $\g_{\mu\nu}(\theta_0)$, denoted $\g'_{\mu\nu}(\theta_0)$, where we switch back from classic index notation to denoting a coordinate change with a prime on the tensor itself.  As previously prescribed, to each instance of $\theta_0$ in the coordinate components of this new metric, add $(\theta-\theta_0)\chi$, denoting the result as $\g'_{\mu\nu}\big(\theta)=\g'_{\mu\nu}(\theta_0+(\theta-\theta_0)\chi\big)$.  It proves convenient to divide $\hP$ and $\hP'$ into two parts,
\begin{align}\label{eq:hP}
\hP&=\frac{1}{2}\int_M d\mathring\mu\,\hT^{\mu\nu}\left.\frac{d}{d\theta}\right|_{\theta_0}\g_{\mu\nu}(\theta)
=\frac{1}{2}\int_Kd\mathring\mu\,\hT^{\mu\nu}\left.\frac{d}{d\theta}\right|_{\theta_0}\g_{\mu\nu}(\theta)
+\frac{1}{2}\int_{\bar{K}}d\mathring\mu\,\hT^{\mu\nu}\left.\frac{d}{d\theta}\right|_{\theta_0}\g_{\mu\nu}(\theta)
=\hP_K+\hP_\barK\;,\\
\label{eq:hPprime}
\hP'&=\frac{1}{2}\int_M d\mathring\mu'\,\hT^{\prime\mu\nu}\left.\frac{d}{d\theta}\right|_{\theta_0}\g'_{\mu\nu}(\theta)
=\frac{1}{2}\int_Kd\mathring\mu'\,\hT^{\prime\mu\nu}\left.\frac{d}{d\theta}\right|_{\theta_0}\g'_{\mu\nu}(\theta)
+\frac{1}{2}\int_{\bar{K}}d\mathring\mu'\,\hT^{\prime\mu\nu}\left.\frac{d}{d\theta}\right|_{\theta_0}\g'_{\mu\nu}(\theta)
=\hP'_K+\hP'_\barK\;,
\end{align}
where $\bar{K}=M\backslash K$.  The integrals over $\bar K$ are restricted to the neighborhood of $K$ where the bump function is nonzero, but in which $\<{\no{\hT^{\mu\nu}}}=0$.  The content of Eq.~(\ref{diffeo}) is that
\begin{align}\label{eq:hPprimeK}
\hP'_K=\hP_K+\int_{\partial K} d\lambda\,n_\mu\hat{T}^{\mu\nu}X_\nu=\hP_K+\hB\;,
\end{align}
where $\hB$ is the boundary term.  By construction, we have $\<{\no{\hP'_{\bar K}}}=\<{\no{\hP_{\bar K}}}=\<{\no{\hB}}=0$, so
\begin{align}\label{eq:hPs}
\<{\no{\hP'}}=\<{\no{\hP'_K}}=\<{\no{\hP_K}}=\<{\no{\hP}}\;;
\end{align}
this is the general version of what we showed for the particular case of Schwarzschild and isotropic coordinates.

What we need for the Cram\'er-Rao bound~(\ref{thur}) are variances, not mean values, and it is in the variances that the problem with reference-state contributions arises.  Equation~(\ref{eq:hP}) gives us
\begin{align}
\begin{split}
\<{(\Delta\hP)^2}&=\<{\no{\hP}\no{\hP}}-\<{\no{\hP}}^2\\
&=\big\langle\!\no{(\hP_K+\hP_{\bar{K}})}\no{(\hP_K+\hP_{\bar{K}})}\!\big\rangle-\<{\no{\hP_K}}^2\\
&=\<{(\Delta\hP_K)^2}+\<{\no{\hP_K}\no{\hP_{\bar{K}}}}+\<{\no{\hP_{\bar{K}}}\no{\hP_K}}+\<{\no{\hP_{\bar{K}}}\no{\hP_{\bar{K}}}}\;,
\label{Pvar}
\end{split}
\end{align}
where $\<{(\Delta\hP_K)^2}=\<{\no{\hP_K}\no{\hP_K}}-\<{\no{\hP_K}}^2$.  We now decompose $\no{\hP_K}$ into its expectation value and a correction,  $\no{\hP_K}=\<{\no{\hP_K}}+\no{\Delta\hP_K}$, and use the fact that $\<{\no{\hP_K}\no{\hP_{\bar{K}}}}=\<{\no{\hP_K}}\<{\no{\hP_{\bar{K}}}}+\<{\no{\Delta\hP_K}\no{\hP_{\bar{K}}}}=\<{\no{\Delta\hP_K}\no{\hP_{\bar{K}}}}$ to write the variance in the form
\begin{align}
\<{(\Delta\hP)^2}
=\<{(\Delta\hP_K)^2}+\<{\no{\Delta\hP_K}\no{\hP_{\bar{K}}}}+\<{\no{\hP_{\bar{K}}}\no{\Delta\hP_K}}+\<{\no{\hP_{\bar{K}}}\no{\hP_{\bar{K}}}}\;.
\end{align}
The final three terms are all reference-state contributions: the middle two terms express correlations between the reference state inside and outside of $K$; the third term is a reference-state contribution from outside the support of the probe's mean stress-energy.

Now, using Eqs.~(\ref{eq:hPprime}) and~(\ref{eq:hPprimeK}) to write $\hP'=\hP_K+\hQ$, with $\hQ=\hB+\hP'_{\bar K}$, the same considerations give us
\begin{align}
\begin{split}
\<{(\Delta\hP')^2}
&=\<{\no{\hP'}\no{\hP'}}-\<{\no{\hP'}}^2\\
&=\big\langle\!\no{(\hP_K+\hQ)}\no{(\hP_K+\hQ)}\!\big\rangle-\<{\no{\hP_K}}^2\\
&=\<{(\Delta\hP_K)^2}+\<{\no{\hP_K}\no{\hQ}}+\<{\no{\hQ}\no{\hP_K}}+\<{\no{\hQ}\no{\hQ}}\\
&=\<{(\Delta\hP_K)^2}+\<{\no{\Delta\hP_K}\no{\hQ}}+\<{\no{\hQ}\no{\Delta\hP_K}}+\<{\no{\hQ}\no{\hQ}}\;.
\end{split}
\label{Primevar}
\end{align}
\end{widetext}
Again, the final three terms are reference-state contributions with
the same sort of interpretation as that given above for
$\<{(\Delta\hP)^2}$, except that now there is a contribution from the
boundary $\partial K$.

If we take advantage of the improved signal-to-noise that comes from
a large mean field, we expect that our probe fields have a sufficiently
substantial mean component that $\<{(\Delta\hP_K)^2}$ makes the dominant
contribution to the variances and thus determines the quantum Cram\'er-Rao bound.
Under these assumptions we conclude that $\<{(\Delta\hP')^2}$ and
$\<{(\Delta\hP)^2}$ are both equal to $\<{(\Delta\hP_K)^2}$ up to
subleading, mean-field-independent terms.

A complementary point of view acknowledges that for a real perturbation,
not one that has been modified by a bump function, the
changes in the reference state do contribute to the quantum Cram\'er-Rao bound.
These changes in the reference state (in some cases, the reference state could
be vacuum, when that can be properly defined) could presumably be used to
detect the perturbation in the absence of a mean field.  Yet we do not know
how to calculate the reference-state contributions in general curved spacetimes,
nor do we know how to measure the corresponding modifications of the reference state.
It would be desirable to determine how to do the necessary measurements, which
might involve particle emission or Casimir-type effects.  What we can say
in the present context is that when we assume that the mean-field terms predominate,
we are finding quantum Cram{\'e}r-Rao bounds on measurements that we do know
how to perform, which involve large mean fields.

\section{Simple examples}
\label{sec:metricestimation}

\subsection{Estimation of Constant Metric Components}

Consider now making measurements in a local inertial frame where the
fiducial metric $\g(0)$ is flat or sufficiently flat for differences to be negligible in
our calculations.  In this case we define a local inertial coordinate system where the
fiducial metric is $\eta_{\mu\nu}$.  Now further suppose that the perturbed metric has variation $\g_{\mu\nu}(\theta)=\eta_{\mu\nu}+\theta\delta_\mu^{\mu_0}\delta_\nu^{\nu_0}$, for some fixed $\mu_0$ and $\nu_0$.
In other words, assume that the parameter of interest is $\theta$, and the fixed local coordinates are such that $\delta\g_{\mu_0\nu_0}=\theta$.  Then the uncertainty relation~(\ref{thur}) becomes
\begin{align}
\langle(\delta\g_{\mu_0\nu_0})^2\rangle
\bigg\langle\bigg(\Delta \int_Kd\mathring\mu\,\hat{T}^{\mu_0\nu_0}\bigg)^{\!2}\,\bigg\rangle
\geq\hbar^2\;,
\label{psmur}
\end{align}
where in this case there is no sum over the repeated indices as we are dealing with a particular metric component specified by the fixed values $\mu_0$ and $\nu_0$.  This inequality is reminiscent of the Unruh uncertainty relation~(\ref{umur}).  Indeed, the same restrictions were needed to derive Eq.~(\ref{umur}) as were used to produce Eq.~(\ref{psmur}).

It should be emphasized that the stress-energy tensor $\hat{T}^{\mu\nu}$ in Eq.~(\ref{psmur}) is for the probe field. It is not the stress-energy tensor for the matter distribution which gives rise to $\g_{\mu\nu}$ via Einstein's field equations.  This is to be expected, as Eq.~(\ref{psmur}) is essentially an uncertainty relation for the probe field: we are estimating the metric with field measurements, so the uncertainty in the field in Eq.~(\ref{psmur}) has been replaced by uncertainty in our estimate of the metric, just as uncertainty in some ``clock'' variable is replaced by uncertainty in a time estimate in the time-energy uncertainty relation~(\ref{teur}).   The uncertainty relation~(\ref{psmur}) can be thought of as the minimum uncertainty achievable when attempting to verify with measurements that the metric takes the Minkowski form in the local coordinate system one has defined.

\subsection{Estimation of Proper Time and Proper Distance}

Suppose we are interested in the proper time as measured by a stationary observer in a perturbed Minkowski spacetime.
Assuming the metric perturbation is compactly supported in space, we might consider the passage of time as measured by an atomic clock at rest within the perturbed region, relative to the passage of time as measured by an atomic clock at rest in flat spacetime outside the perturbed region.  Both can be considered proper time, but the latter is also equivalent to our Minkowskian coordinate time.  The proper time in our locally defined inertial frame is related to the coordinate time by
\begin{align}
\tau=\int dt\,\sqrt{-\g_{00}}\;.
\end{align}

In these coordinates, then, we are interested in a metric perturbation of the form $\g_{00}^{(s)}=\eta_{00}+sa(x)$,
where $a$ is a smooth function of the coordinate 4-position $x$.  Using the approximation\begin{align}
\langle[\delta f(X)]^2\rangle\simeq
\left(\left.\frac{df}{dX}\right|_{X =\langle X\rangle}\right)^{\!2}
\langle(\delta X)^2\rangle\;,
\end{align}
the uncertainty in the metric is related to the uncertainty in the proper time by
\begin{align}
\langle(\delta s)^2\rangle=4\left(\int dt\,a(x)\right)^{\!-2}\!\!\langle(\delta \tau)^2\rangle\;.
\end{align}
Then the relation~(\ref{eq:qcrb}) becomes
\begin{align}
\left(\int dt\,a(x)\right)^{\!-2}\!\!
\langle(\delta \tau)^2\rangle
\left\langle\!\left(\Delta \int d^4x\, \hat{T}^{00}(x)a(x)\right)^{\!2}\right\rangle
\geq\frac{\hbar^2}{4}\;.
\label{tmur3}
\end{align}

Now we assume that the effective spatial volume of the confined probe field is small enough that the spacetime perturbation can be considered spatially uniform throughout.  Then, recognizing that the integral of the energy density over the spatial component of the four-volume is the Hamiltonian, we have
\begin{align}
\int d^4x\,\hat{T}^{00}(x)a(t)=\int dt\,\hat{H}(t)a(t)\;.
\end{align}
Further assuming a time-independent Hamiltonian, the time-integrals cancel and the uncertainty relation
reduces to
\begin{align}
\langle(\delta \tau)^2\rangle\langle(\Delta \hat{H})^2\rangle\geq\frac{\hbar^2}{4}\;.
\end{align}
This demonstrates that the standard time-energy uncertainty relation is a special case of the metric uncertainty relation~(\ref{eq:qcrb}).

Using a similar argument, one can derive a corresponding uncertainty relation for proper distance $X$:
\begin{align}
\langle(\delta X)^2\rangle\langle(\Delta \hat{P}_X)^2\rangle\geq\frac{\hbar^2}{4}\;.
\end{align}
This is the parametric version of the Heisenberg uncertainty relation, where $\hat{P}_X$ is the momentum in the direction of the displacement. It demonstrates the consistency between the metric uncertainty relation~(\ref{eq:qcrb}) and the earlier work on parameter-based uncertainty relations for the Lorentz group in flat spacetime~\cite{Braunstein:1996}.

\section{Quantum-limited gravitational-wave detection}
\label{sec:gw}

We now consider estimating the amplitude of a gravitational wave.  To a good approximation, a gravitational wave can be modeled by a small perturbation of Minkowski spacetime satisfying the linearized Einstein
equations.  Thus we write
\begin{align}
\g_{\mu\nu}=\eta_{\mu\nu}+h_{\mu\nu}\;,
\end{align}
where for a plane-fronted, parallel-propagating wave, linearly polarized along the $x$ and $y$ axes and cast in the transverse-traceless gauge \cite{mtw}, the nonvanishing components of the metric perturbation are
\begin{equation}
\label{gravwave}
h_{xx}=-h_{yy}=A_+f(z-t)\;.
\end{equation}
Physical solutions have a suitably localized envelope along the propagation direction, which can be approximated by a compactly supported function of $z-t$. Physical solutions also are not exactly plane-fronted, but rather are confined in the directions transverse to the propagation direction or, as in the case of astrophysical sources, have spherical wave fronts.  We assume for our analysis that any physical deviations from a plane-fronted wave are negligible on the spatial scales of our probe field.  Assuming that the gravitational-wave detector, which might be the electromagnetic field confined within a laser-powered interferometer, is compactly supported in space, the intersection with the detector's support is  compactly supported in spacetime.  The volume integrals of concern to us are therefore well defined.

For simplicity, however, we analyze broadband detection of a gravitational wave, i.e., detection that is essentially instantaneous compared to the scale of variation (period or wavelength) of the gravitational wave.  What this means is that the probe field's support is sufficiently confined spatially and temporally relative to the gravitational wave's envelope and wavelength that within the probe's window of observation, the gravitational wave is well approximated by a constant:
\begin{equation}
h_{xx}=-h_{yy}\simeq A\;.
\end{equation}
For consistency with this assumption and to ensure that the relevant volume integrals remain well defined, we assume a finite duration of detection.  Since our perturbed metric happens to be in Gaussian normal coordinates (i.e., $\g_{00}=-1$ and $\g_{0i}=0$), the resulting coordinate bounds of integration are independent of the perturbation.  Our assumptions amount to saying that the probe field is to be turned on and off, i.e., emitted and absorbed, within a compact spatial region in such a way that it senses an essentially instantaneous amplitude of the gravitational wave over this compact spatial region.

The generator~(\ref{Ptheta}) of changes in the probe field is
\begin{align}
\label{Pwave}
\begin{split}
\hP
&= \frac{1}{2}\int d^4x\,\hT^{\mu\nu}\left.\frac{d}{dA}\right|_0\g_{\mu\nu}(A)\\
&= \frac{1}{2}\int d^4x \left(\hT^{xx}\left.\frac{d}{dA}\right|_0 \g_{xx}(A)
+\hT^{yy}\left.\frac{d}{dA}\right|_0\g_{yy}(A)\right)\\
&= \frac{1}{2}\int d^4x\,(\hT^{xx}-\hT^{yy})\;,
\end{split}
\end{align}
where we have assumed the fiducial amplitude is zero (i.e., perturbation
about flat spacetime).  The domain of integration encloses the finite
extent of the probe mean~field.

We now take the probe field to be the electromagnetic field, having
a large mean field that is turned on and off, as we have discussed.
We assume that the domain of integration in
Eq.~(\ref{Pwave}) is large, both temporally and spatially, compared to
the scales of variation (periods and wavelengths) of the mean
electromagnetic field.  We could regard the probe electromagnetic
field as being confined within a laser-powered interferometer, as in
the LIGO detectors~\cite{LIGO2015a,LIGO2016a,LIGO2016b}, but there is
no need to specialize to this particular field configuration.
Instead, we let the probe be a free electromagnetic field: the field
is turned on, receives an imprint from the gravitational wave as it
propagates freely through the gravitational wave, and is then turned
off.  Recall that the Cram\'er-Rao bound optimizes over all
measurements we could make on the probe field, so we do not have to
specify what measurement is used to read out the imprint of the
gravitational wave on the electromagnetic field, although we will have
something to say about this as we proceed.  This approach allows us to
use the free electromagnetic field and the free-field commutators.  In
this approach, it is clear that we do not find a ``standard quantum
limit'' that is enforced by back-action forces that act on masses that
confine the field, because there are no such masses.

Notice that if we did regard the field as being in an interferometric configuration, there would need to be beam splitters and mirrors to split, confine, and recombine the field.  To neglect back-action and thus to be consistent with the present calculation, we could make these optical elements sufficiently massive that they are unaffected by the field's back-action radiation-pressure noise and thus move on geodesics.  All of this is consistent with the now well-established result that there is no back-action-enforced ``standard quantum limit'' that fundamentally limits interferometric gravitational-wave detectors.  The absence of a back-action-enforced fundamental limit for interferometric detectors follows from a substantial body of work on specialized, back-action-evading designs for laser-interferometer gravitational-wave detectors~\cite{unruh,kimble,khalili} and from general analyses of quantum limits on the detection of waveforms~\cite{tsang,caves}.

For the electromagnetic field, the diagonal components of the stress tensor are
\begin{equation}\label{eq:Tjj}
\hT^{jj}=\frac{1}{8\pi}\big(\hE_x^2+\hE_y^2+\hE_z^2+\hB_x^2+\hB_y^2+\hB_z^2\big)
-\frac{1}{4\pi}\big(\hE_j^2+\hB_j^2\big)\;,
\end{equation}
where $\hE_j^2$ and $\hB_j^2$ are normally ordered and we use cgs Gaussian units with $c=1$.  The local generator of changes in the field due to the gravitational wave is
\begin{align}
\begin{split}
\frac{1}{2}(\hT^{xx}-\hT^{yy})&=\frac{1}{8\pi}(\hE_y^2-\hE_x^2+\hB_y^2-\hB_x^2)\\
&=\frac{1}{8\pi}\sum_\sigma r_\sigma \no{\hf_\sigma^2}\;.
\end{split}
\end{align}
Here we let $\hf_1=\hE_y$, $\hf_2=\hE_x$, $\hf_3=\hB_y$, $\hf_4=\hB_x$ and $r_1=r_3=1$, $r_2=r_4=-1$.  Beginning with the last form, we indicate normal ordering explicitly where it is needed.  The generator~(\ref{Pwave}) becomes
\begin{align}
\no{\hP}=\frac{1}{8\pi}\sum_\sigma r_\sigma\int d^4x\,\no{\hf_\sigma^2}\;.
\end{align}

Now we express the electric and magnetic fields as a sum of a mean field and
field fluctuations, defined as the deviation from the mean:
\begin{align}
\hf_\sigma(\vect x,t)=\<{\hf_\sigma(\vect x,t)}+\Delta\hf_\sigma(\vect x,t)\;.
\end{align}
This puts the generator in the form
\begin{align}\label{eq:noP}
\no{\hP}=P+\Delta\hX_1+\no{\hF}\;,
\end{align}
where
\begin{align}
P&=\frac{1}{8\pi}\sum_\sigma r_\sigma\int d^4x\,\<{\hf_\sigma}^2\;,\\
\hX_1&=\frac{1}{4\pi}\sum_\sigma r_\sigma\int d^4x\,\<{\hf_\sigma}\hf_\sigma\;,\\
\no{\hF}&=\frac{1}{8\pi}\sum_\sigma r_\sigma\int d^4x\,\no{(\Delta\hf_\sigma)^2}\;.\label{eq:Fhat}
\end{align}
Notice that we do not need to normal order $\hX_1$ because it is linear in field operators.

Note that by our formalism, the quantum fields in
Eqs.~(\ref{eq:Tjj})--(\ref{eq:Fhat}) need only be evaluated in the
fiducial spacetime, which in the present case is flat.  Therefore, the
vacuum state is unambiguous, and the splitting of the field operators
into positive- and negative-frequency parts and the use of normal
ordering are appropriate and well defined.  (For a discussion of the
issues arising in curved spacetime, see Sec.~1 of \cite{bfv}.)
Thus we have
\begin{align}\label{eq:DeltahatFsigmasquared}
\begin{split}
\no{\Delta\hf_\sigma^2}
&=\no{[\Delta\hf_\sigma^{(+)}+\Delta\hf_\sigma^{(-)}]^2}{}\\
&={}2\Delta\hf_\sigma^{(-)}\Delta\hf_\sigma^{(+)}
+\Delta\hf_\sigma^{(+)2}+\Delta\hf_\sigma^{(-)2}\;.
\end{split}
\end{align}

The free-field commutators and vacuum correlators that we need
are summarized in Appendix~\ref{sec:commcorr}.

Our separation of the field operators into a
mean field plus field fluctuations is different from our treatment in
Sec.~\ref{sec:compact}, where we separated the stress-energy, which is
generally quadratic in field operators, into its mean and its
fluctuation about the mean.  To identify the mean-field-independent
contributions to the variance, we view the state as being obtained
from a zero-mean-field state by a displacement operator $D$, which
is generated by a linear function of the fields.  This operator is
determined by requiring that $D^{\dagger}f_{\sigma}D = f_{\sigma}+\<{f_{\sigma}}$.
The displacement parameter is the mean field.  For the present purposes,
the zero-mean-field state is the reference state and is considered fixed.
One example is where this reference state is the vacuum state.  Our
initial arguments apply to all zero-mean-field reference states,
except where noted otherwise, and we eventually get to the case of
a squeezed-vacuum state as the reference state that provides
optimal sensitivity under the assumptions we make.  Our main
conclusions are aimed at the case where the displacement is large,
in which case we only keep the terms that are leading order in the
displacement.

The expectation value of the generator~(\ref{eq:noP}) is
\begin{align}
\<{\no{\hP}}=P+\<{\no{\hF}}\;.
\end{align}
  Using
\begin{align}\label{eq:noPdecomposition}
\begin{split}
\no{\hP}\no{\hP}&=P^2+2P\no{\hF}+(\Delta\hX_1)^2+\no{\hF}\no{\hF}\\
&\quad+2P\Delta\hX_1+\Delta\hX_1\no{\hF}+\no{\hF}\Delta\hX_1\;,
\end{split}
\end{align}
we have
\begin{align}\label{eq:noPsquared}
\begin{split}
\<{\no{\hP}\no{\hP}}&=P^2+2P\<{\no{\hF}}+\<{(\Delta\hX_1)^2}+\<{\no{\hF}\no{\hF}}\\
&=\<{\no{\hP}}^2+\<{(\Delta\hX_1)^2}+\<{\no{\hF}\no{\hF}}-\<{\no{\hF}}^2\;.
\end{split}
\end{align}
Here we assume that the odd moments of the reference (zero-mean-field)
state are zero, which is the case if the reference state is Gaussian or
is invariant under parity and time reversal.

Rewriting Eq.~(\ref{eq:noPsquared}) in terms of the variance, we~get
\begin{align}
\<{(\Delta\hP)^2}&=\<{\no{\hP}\no{\hP}}-\<{\no{\hP}}^2=\<{(\Delta\hX_1)^2}+\<{(\Delta\hF)^2}\;.
\end{align}
We subsume the normal ordering into the definition of the definition
of the variance $(\Delta\hF)^{2}$ when using this notation.  The term
$\<{(\Delta\hF)^2}$ is mean field independent, so for large mean field,
we can drop it.  Before doing so, however, it is worth taking a
closer look at the mean-field-independent contributions.  When we put the
right-hand side of Eq.~(\ref{eq:DeltahatFsigmasquared}) into the spacetime
integral~(\ref{eq:Fhat}) to get $\no{\hF}$, we can expand the field operators
in the last two terms of Eq.~(\ref{eq:DeltahatFsigmasquared}) into
integrals over the wave vectors of free-field plane-wave modes, as in
Appendix~\ref{sec:commcorr}.  Performing the spacetime integral first, the
amplitudes for a pair of wave vectors, $\vect k$ and $\vect k'$, average to
nearly zero, except for field modes whose period and wavelength are as large
or larger than the temporal and spatial extent of the region of integration.
Realistic measurement devices such as laser interferometers are neither
designed for nor capable of detecting such low-frequency photons.
If we neglect these essentially DC contributions, we are left with
\begin{align}\label{eq:normhatF}
\no{\hF}
=\frac{1}{4\pi}\sum_\sigma r_\sigma\int d^4x\,\Delta\hf_\sigma^{(-)}\Delta\hf_\sigma^{(+)}\;,
\end{align}
where the equals sign now assumes that we have omitted the DC contributions.  The corresponding
variance is
\begin{widetext}
\begin{align}\label{eq:varianceF}
\begin{split}
\<{(\Delta\hF)^2}=\<{\no{\hF}\no{\hF}}-\<{\no{\hF}}^2
=\frac{1}{16\pi^2}\sum_{\sigma,\sigma'}r_\sigma r_{\sigma'}\!\int d^4x\,d^4x'
\Big(
&\big\langle\Delta\hf_\sigma^{(-)}(\vect x,t)\Delta\hf_\sigma^{(+)}(\vect x,t)
\Delta\hf_{\sigma'}^{(-)}(\vect x',t')\Delta\hf_{\sigma'}^{(+)}(\vect x',t')\big\rangle\\
&-\big\langle\Delta\hf_\sigma^{(-)}(\vect x,t)\Delta\hf_\sigma^{(+)}(\vect x,t)\big\rangle
\big\langle\Delta\hf_{\sigma'}^{(-)}(\vect x',t')\Delta\hf_{\sigma'}^{(+)}(\vect x',t')\big\rangle
\Big)\;.
\end{split}
\end{align}
\end{widetext}

If the electromagnetic field is excited into a coherent state, where the
field fluctuations are those of vacuum, both $\<{\no{\hF}}$ and
$\<{(\Delta\hF)^2}$, as calculated from Eqs.~(\ref{eq:normhatF})
and~(\ref{eq:varianceF}) vanish.  For coherent states, the only
mean-field-independent contributions to the total variance of $\hF$
come from the DC terms discarded in going from
Eq.~(\ref{eq:DeltahatFsigmasquared}) to Eq.~(\ref{eq:normhatF}).

If the field fluctuations are redistributed relative to vacuum, as in the
squeezed state discussed below, the terms of $\<{(\Delta\hF)^2}$
given in Eq.~(\ref{eq:varianceF}) make the dominant mean-field-independent
contribution, expressing the fact that these nonvacuum field fluctuations are
affected by the presence of a gravitational wave and can be used to
detect the wave. For sufficiently large mean field, these
mean-field-independent contributions are small compared to
$\<{(\Delta\hX)_1^2}$, leaving us with
\begin{align}
\<{(\Delta\hP)^2}=\<{(\Delta\hX_1)^2}\;,
\end{align}
as we assume henceforth.

We can now summarize our results by saying that the Cram\'er-Rao bound~(\ref{eq:qcrb}) on the estimate of the gravitational-wave amplitude~$A$ is
\begin{align}\label{eq:qcrbA}
\langle(\delta \tilde{A})^2\rangle\langle(\Delta \hX_1)^2\rangle\geq\frac{\hbar^2}{4}\;.
\end{align}
We could stop here, having confirmed the valuable lesson, generic to
Cram\'er-Rao bounds, that precise determination of the
gravitational-wave amplitude requires that the observable~$\hX_1$,
which in the presence of a large mean field generates the change in
the probe-field state, be as uncertain as possible.  In this
case, however, we can say considerably~more.

Since $\hX_1$ is linear in the fields, one can find an observable
$\hX_2$, conjugate to $\hX_1$ and also linear in the fields, which
is the observable one should measure to effect the precise determination
of~$A$. The commutator of $\hX_1$ and $\hX_2$ is
\begin{align}\label{eq:QUADcomm}
\big[\hX_1,\hX_2\big]=i\hbar C\;,
\end{align}
where the real constant $C$ is to be determined (we can make $C$ positive by, say, changing the sign of $\hX_2$).  The commutator implies a Heisenberg uncertainty relation, precisely analogous to the position-momentum uncertainty relation~(\ref{eq:heisenberg}),
\begin{align}
	\langle(\Delta \hX_1)^2\rangle\langle(\Delta \hX_2)^2\rangle\geq\frac{\hbar^2}{4}C^2\;.
	\label{eq:HeisenbergQUAD}
\end{align}
To put these two observables on the same footing relative to vacuum, we require that
\begin{align}\label{eq:QUAD2m}
\big\langle0\big|\hX_1^2\big|0\big\rangle=\big\langle0\big|\hX_2^2\big|0\big\rangle=\frac{\hbar}{2}C\;.
\end{align}
The observables $\hX_1$ and $\hX_2$ are generalized
quadrature components~\cite{caves80,cavesschu85,schucaves85} for
the single field mode that is determined with respect to the Minkowski
vacuum by the mean field according to the definition of $\hX_{1}$,
with their vacuum level of noise given by $\hbar C/2$.  We calculate
$\hX_{2}$ explicitly below after restricting to the case of a plane wave.

Equation~(\ref{mainBFVresult}) specifies the response of $\hX_2$ to
the gravitational wave,
\begin{align}
\frac{d\hX_2}{dA} =\frac{i}{\hbar}\big[\hX_2,\hX_1\big]=C\;.
\end{align}
Linear-response analysis gives the variance of an estimate of $A$ based
on a measurement of $\hX_2$,
\begin{align}\label{eq:qcrbAQUAD}
\<{(\delta\tilde A)^2}
=\frac{\<{(\Delta\hX_2)^2}}{\big|d\<{\hX_2}/dA\big|^2}
=\frac{\<{(\Delta\hX_2)^2}}{C^2}
\ge\frac{\hbar^2}{4}\frac{1}{\<{(\Delta\hX_1)^2}}\;,
\end{align}
matching the Cram\'er-Rao bound~(\ref{eq:qcrbA}).  If the probe field is placed in a minimum-uncertainty state relative to the uncertainty relation~(\ref{eq:HeisenbergQUAD}), the bound~(\ref{eq:qcrbAQUAD}) is saturated, and it is particularly useful to write the variance of the estimate as
\begin{align}\label{eq:varAQUAD}
\<{(\delta\tilde A)^2}
=\frac{\<{(\Delta\hX_2)^2}}{C^2}
=\frac{\hbar}{2C}\sqrt{\frac{\<{(\Delta\hX_2)^2}}{\<{(\Delta\hX_1)^2}}}\;.
\end{align}
We stress that Eq.~(\ref{eq:varAQUAD}) is not a general expression for the Cram\'er-Rao bound, but rather is the form the bound assumes for minimum-uncertainty states relative to the uncertainty relation~(\ref{eq:HeisenbergQUAD}).

The physical content here is that if the field is excited into a coherent state, the uncertainties in the quadrature components are equal, and the variance of the estimate of $A$, equal to $\hbar/2C$, is set by the vacuum-level noise in $\hX_1$ and $\hX_2$.  To achieve a sensitivity better than $\hbar/2C$, one should squeeze the vacuum so that the uncertainty in $\hX_2$ decreases and the uncertainty in $\hX_1$ increases, as in the original proposal for decreasing shot noise in a laser-interferometer gravitational-wave detector by using squeezed light~\cite{caves81}, a proposal that has been implemented in large-scale laser-interferometer detectors~\cite{LIGO11,LIGO13} and might be incorporated into Advanced LIGO~\cite{Miller2015}.

There is one task remaining, quite an important one, and that is to evaluate the constant~$C$.  To do that, we specialize a bit, to the case where the mean probe field is that of a nearly plane wave propagating in the $x$ direction and linearly polarized along the $y$ axis.  We do not need to assume that this wave is close to monochromatic, but we do assume that the transverse extent of the wave is much larger than the wave's typical wavelengths.  We neglect the small corrections to a plane wave due to the finite transverse extent.  With these assumptions, we have $\<{\hE_x}=\<{\hE_z}=\<{\hB_x}=\<{\hB_y}=0$ and
\begin{align}\label{eq:Ey}
\<{\hE_y}=\<{\hB_z}=E_1(\vect x,t)\;,\phantom{\int}
\end{align}
where the (real) waveform $E_1(\vect x,t)$ is mainly a function of $x-t$ and only a weak function of $y$ and $z$.  With these assumptions, we have
\begin{align}\label{eq:hX1}
\hX_1=\frac{1}{4\pi}\int d^4x\,E_1(\vect x,t)\hE_y(\vect x,t)\;.
\end{align}

It is useful to divide $E_1$ into positive- and negative-frequency parts and to write these in terms of the Fourier transform,
\begin{widetext}
\begin{align}
\label{eq:E1}
E_1(\vect x,t)&=E_1^{(+)}(\vect x,t)+E_1^{(-)}(\vect x,t)\;,\\
\label{eq:E1plus}
E_1^{(+)}(\vect x,t)&=E_1^{(-)*}(\vect x,t)
=i\sum_\sigma\int\dthreek\,\sqrt{2\pi\hbar\omega}\,\alpha_{1;\ks}\vect{e}_\ks\cdot\vect e_y
e^{i(\vect k\cdot\vect x-\omega t)}
=i\int\dthreek\,\sqrt{2\pi\hbar\omega}\,\alpha_{1,\vect k}e^{i(\vect k\cdot\vect x-\omega t)}\;,
\end{align}
where $\alpha_{1;\ks}=\<{a_\ks}$.  The assumption of a nearly plane wave propagating in the $x$ direction is that $\alpha_{1;\ks}$ has substantial support only for $\vect k$ pointing nearly along the $x$ direction, with linear polarization nearly along the $y$ direction, in which case we drop the polarization index and write it as $\alpha_{1,\vect k}$ (formally, we might write $\alpha_{1;\ks}=\delta_{\sigma y}\alpha_{1,\vect k}$); this leads to the final form in Eq.~(\ref{eq:E1plus}).

We assume that $\hX_2$ looks the same as $\hX_1$,
\begin{align}\label{eq:hX2}
\hX_2=\frac{1}{4\pi}\int d^4x\,E_2(\vect x,t)\hE_y(\vect x,t)\;,
\end{align}
but with a different (real) waveform $E_2$, which is also a nearly plane wave propagating in the $x$ direction, with linear polarization nearly along the $y$ direction,
\begin{align}
\label{eq:E2}
E_2(\vect x,t)&=E_2^{(+)}(\vect x,t)+E_2^{(-)}(\vect x,t)\;,\\
\label{eq:E2plus}
E_2^{(+)}(\vect x,t)&=E_2^{(-)*}(\vect x,t)
=i\sum_\sigma\int\dthreek\,\sqrt{2\pi\hbar\omega}\,\alpha_{2;\ks}\vect{e}_\ks\cdot\vect e_y e^{i(\vect k\cdot\vect x-\omega t)}
=i\int\dthreek\,\sqrt{2\pi\hbar\omega}\,\alpha_{2,\vect k}e^{i(\vect k\cdot\vect x-\omega t)}\;.
\end{align}
Again we understand that $\alpha_{2,\vect k}$ has support only for $\vect k$ close to the $x$ direction and corresponds to linear polarization nearly along the $y$ direction ($\alpha_{2;\ks}=\delta_{\sigma y}\alpha_{2,\vect k}$).  The following calculations show that the above assumption is warranted.  Specifically, we find in Eq.~(\ref{eq:alpha21}) that $E_2$ can be obtained from $E_1$ by a $90^\circ$ phase shift of every monochromatic mode that contributes to $E_1$, as one might expect for a broadband version of conjugate quadrature components.

Notice that in the expressions~(\ref{eq:hX1}) and~(\ref{eq:hX2}) for $\hX_1$ and $\hX_2$, we can extend the spatial integrals over all of space because the waveforms $E_1(\vect x,t)$ and $E_2(\vect x,t)$ are zero outside the original domain of spatial integration.

To determine $C$, we use the field commutators and vacuum correlators of Appendix~\ref{sec:commcorr} [see Eqs.~(\ref{eq:EorBcomm}) and (\ref{eq:EorBcorr})] to find the commutator~(\ref{eq:QUADcomm}) and the second moments~(\ref{eq:QUAD2m}):
\begin{align}
\begin{split}\label{eq:QUADcomm2}
\big[\hX_1,\hX_2\big]
&=\frac{1}{16\pi^2}\int d^4x\,d^4x'\,E_1(\vect x,t)E_2(\vect x',t')\big[\hE_y(\vect x,t),\hE_y(\vect x',t')\big]\\
&=\frac{i\hbar}{16\pi^2}\int d^4x\,d^4x'\,E_1(\vect x,t)E_2(\vect x',t')
\left(\frac{\partial^2}{\partial t^2}-\frac{\partial^2}{\partial y^2}\right)
G(\vect x-\vect x',t-t')\;,
\end{split}\\
\begin{split}\label{eq:QUAD2m2}
\big\langle0\big|\hX_a^2\big|0\big\rangle
&=\frac{1}{16\pi^2}\int d^4x\,d^4x'\,E_a(\vect x,t)E_a(\vect x',t')
\frac12\big\langle0\big|[\hE_y(\vect x,t)\hE_y(\vect x',t')+\hE_y(\vect x',t')\hE_y(\vect x,t)]\big|0\big\rangle\\
&=\frac{\hbar}{16\pi^3}\int d^4x\,d^4x'\,E_a(\vect x,t)E_a(\vect x',t')
\left(-\frac{\partial^2}{\partial t^2}+\frac{\partial^2}{\partial y^2}\right)
D(\vect x-\vect x',t-t')\;,\quad a=1,2.
\end{split}
\end{align}
Here $G(\vect x,t)$, the difference between retarded and advanced Green functions, is defined in Eq.~(\ref{eq:G}), and $D(\vect x,t)$, the principal value of the inverse of the invariant interval, is defined in Eq.~(\ref{eq:D}).

In Eqs.~(\ref{eq:QUADcomm2}) and~(\ref{eq:QUAD2m2}), we can integrate by parts twice on the $y$ derivatives.  The boundary terms vanish because we can take the boundary of the region of integration to be outside the spatial extent of the waveforms $E_1$ and $E_2$, and we can neglect the resulting integrals because $E_1$ and $E_2$ are weak functions of $y$.  The upshot is that we can omit the $y$ derivatives in Eqs.~(\ref{eq:QUADcomm2}) and~(\ref{eq:QUAD2m2}).  Using Eqs.~(\ref{eq:G}) and~(\ref{eq:D}) to start getting back into the Fourier domain, we have
\begin{align}
\big[\hX_1,\hX_2\big]
&=\frac{i\hbar}{4\pi}\,\mbox{Im}\int\dthreek\,\omega
\left(\int dt\,e^{-i\omega t}\int d^3x\,E_1(\vect x,t)e^{i\vect{k}\cdot\vect x}\right)
\left(\int dt'\,e^{i\omega t'}\int d^3x'\,E_2(\vect x',t')e^{-i\vect{k}\cdot\vect x'}\right)
\;,\\
\big\langle0\big|\hX_a^2\big|0\big\rangle
&=\frac{\hbar}{8\pi}\,\mbox{Re}
\int\dthreek\,\omega
\left(\int dt\,e^{-i\omega t}\int d^3x\,E_a(\vect x,t)e^{i\vect{k}\cdot\vect x}\right)
\left(\int dt'\,e^{i\omega t'}\int d^3x'\,E_a(\vect x',t')e^{-i\vect{k}\cdot\vect x'}\right)\;,
\quad a=1,2.
\end{align}
The spatial Fourier transforms are
\begin{align}
e^{-i\omega t}\int d^3x\,E_a(\vect x,t)e^{i\vect{k}\cdot\vect x}
&=\sqrt{2\pi\hbar\omega}\big(\mathord{-}i\alpha_{a,\vect k}^*+i\alpha_{a,-\vect k}e^{-2i\omega t}\big)\;.
\end{align}
\end{widetext}
The counter-rotating terms average to nearly zero in the temporal integrals, so we discard them and obtain
\begin{align}
\int dt\,e^{-i\omega t}\int d^3x\,E_a(\vect x,t)e^{i\vect{k}\cdot\vect x}
=-i\tau\sqrt{2\pi\hbar\omega}\,\alpha_{a,\vect k}^*\;,
\end{align}
where $\tau$ is the time interval over which the mean field is turned on.  Our final results for the commutator and vacuum second moment are
\begin{align}
\big[\hX_1,\hX_2\big]
&=i\hbar\!\left(\frac12\hbar\int\dthreek\,(\omega\tau)^2
\mbox{Im}\big(\alpha_{1,\vect k}^*\alpha_{2,\vect k}\big)\right)\;,\\
\big\langle0\big|\hX_a^2\big|0\big\rangle
&=\frac{\hbar}{2}\!\left(\frac12\hbar\int\dthreek\,(\omega\tau)^2|\alpha_{a,\vect k}|^2\right)\;,
\quad a=1,2.
\end{align}
A glance at Eqs.~(\ref{eq:QUADcomm}) and~(\ref{eq:QUAD2m}) shows that the quantities in large parentheses are all equal to $C$.  Since we want this to be true whatever the probe waveform is, we must have
\begin{align}\label{eq:alpha21}
\alpha_{2,\vect k}=i\alpha_{1,\vect k}\;;
\end{align}
i.e., as promised, $E_2$ is obtained from $E_1$ by a $90^\circ$ phase shift of every monochromatic mode that contributes to $E_1$.  Finally, we obtain
\begin{align}\label{eq:C}
C=\frac12\hbar\int\dthreek\,(\omega\tau)^2|\alpha_{1,\vect k}|^2\;.
\end{align}

Consider now a nearly monochromatic mean field with wave vector $\vect k=\omega\vect e_x$.  The vacuum level of noise in the estimate of the gravitational-wave amplitude is $\hbar/2C=1/(\omega\tau)^2\bar n$, where $\bar n$ is the number of photons carried by the mean field.  We have $E_1^{(+)}\propto\alpha_2 e^{i\omega(x-t)}$ and $E_2^{(+)}\propto i\alpha_2 e^{i\omega(x-t)}$; writing $\alpha_2=|\alpha_2|e^{i\phi}$, we have $E_1\propto\alpha_2\cos[\omega(t-x)-\phi]$ and $E_2\propto\alpha_2\sin[\omega(t-x)-\phi]$.  Thus $\hX_1$ and $\hX_2$ are proportional to the standard quadrature components for a monochromatic field mode.

In the case of a nearly monochromatic mean field, the quantum-limited sensitivity~(\ref{eq:varAQUAD}) for detecting a gravitational-wave amplitude becomes
\begin{align}\label{eq:varAQUADmono}
\<{(\delta\tilde A)^2}
=\frac{1}{(\omega\tau)^2\bar n}\sqrt{\frac{\<{(\Delta\hX_2)^2}}{\<{(\Delta\hX_1)^2}}}\;.
\end{align}
If the field is excited into a nearly monochromatic coherent state, the quadrature components have equal, vacuum-level uncertainties, $\bar n=\<{\hn}=\<{(\Delta\hn)^2}$ is the expectation value and the variance in the number of photons, and the sensitivity is shot-noise-limited, i.e., $\<{(\delta\tilde A)^2}^{1/2}=1/\omega\tau\<{\hn}^{1/2}$.  This result has a physically intuitive interpretation.  The gravitational wave changes the coordinate speed of light in the $x$ direction by $-A/2$, leading to a phase shift $\delta\phi=(\omega\tau)A/2$; the shot-noise limit on detecting the gravitational-wave amplitude translates to $\<{(\delta \phi)^2}\<{(\Delta \hn)^2}=\frac14$, which is the conventional uncertainty-principle bound on phase and photon number.  To do better than shot noise, one can squeeze the $\hX_2$ quadrature, reducing its uncertainty while increasing the uncertainty in the $\hX_2$ quadrature.

A bonus of our approach is that Eq.~(\ref{eq:varAQUAD}) gives us the quantum limit on detecting a gravitational wave using a nearly plane-wave, but broadband probe field:
\begin{align}
\<{(\delta\tilde A)^2}
=\left(\int\dthreek\,(\omega\tau)^2|\alpha_{1,\vect k}|^2\right)^{-1}
\sqrt{\frac{\<{(\Delta\hX_2)^2}}{\<{(\Delta\hX_1)^2}}}\;.
\end{align}
Comparison to the monochromatic sensitivity~(\ref{eq:varAQUADmono}) shows that the way to generalize $(\omega\tau)^2\bar n$ to a broadband mean field is to integrate over contributions from all the monochromatic modes.  If the field is excited into a broadband coherent state, the quantum-limited sensitivity is given by a sort of generalized shot noise quantified by this frequency-weighted integration over mean photon numbers in the monochromatic modes.

To do better than shot-noise-limited sensitivity, one should put the appropriate field mode into a squeezed state.  We can write down the required squeezed state by noting that for a nearly plane wave, the field quadratures take the form $\hX_1=\sqrt{\hbar C/2}\,(\hb+\hb^\dagger)$ and $\hX_2=\sqrt{\hbar C/2}\,(-i\hb+i\hb^\dagger)$ (see Appendix~\ref{sec:commcorr}), where
\begin{align}
\hb=\frac{\tau}{\sqrt{2\hbar C}}\int\dthreek\,\hbar\omega\,\alpha^*_{1,\vect k}\ha_{\vect ky}
\end{align}
and $\hb^\dagger$ satisfy the canonical bosonic commutation relation, $[\hb,\hb^\dagger]=1$.  This means that the desired minimum-uncertainty state is the squeezed state
\begin{align}\label{eq:ss}
e^{\mu\hb^\dagger-\mu^*\hb}\exp\!\bigg(\frac12 r\big[(\hb^\dagger)^2-\hb^2\big]\bigg)|0\rangle\;,
\end{align}
where $\mu$ and $r$ are real, with $\mu$ chosen to give the assumed mean field $E_1(\vect x,t)$.  Indeed, one can see that this state has
\begin{align}
\<{a_{\vect k y}}=\frac{\mu\tau}{\sqrt{2\hbar C}}\,\hbar\omega\,\alpha_{1,\vect k}\;,
\end{align}
so consistency requires that $\mu=\sqrt{2\hbar C}/\hbar\omega\tau$.  The squeezed state~(\ref{eq:ss}) has $\<{\hb}=\mu$, $\<{\hX_1}=\sqrt{2\hbar C}\mu=2C/\omega\tau$, $\<{\hX_2}=0$, and
\begin{align}
\begin{split}
\<{(\Delta\hX_1)^2}&=\frac{\hbar C}{2}e^{2r}\;,\\
\<{(\Delta\hX_2)^2}&=\frac{\hbar C}{2}e^{-2r}\;.
\end{split}
\end{align}
It thus beats shot-noise-limited sensitivity by a factor of $e^{-2r}$.

\section{Other applications}
\label{sec:otherapp}

We now consider briefly a few of the many other applications of our formalism.

Cosmology is one field where accurate measurement of gravitational parameters is of obvious interest.  For a simple example, consider the spatially closed Friedmann-Lemaitre-Robertson-Walker spacetime. This is a universe filled with a uniform density of matter, e.g., ``galaxies,'' and radiation.  At any instant in time, in the comoving frame of the galaxies, the universe looks the same everywhere (homogeneous) and in all directions (isotropic). The metric for this universe is given by~\cite{mtw}
\begin{align}
ds^2=-dt^2+a^2(t)\!\left[d\chi^2+\sin^2\!\chi\left(d\theta^2+\sin^2\!\theta\,d\phi^2\right)\right]\;,
\end{align}
where $t$ is the proper time of an observer comoving with any of the galaxies. The spatial coordinates $\chi,\theta,\phi$ describe homogeneous and isotropic three-spheres of constant proper time $t$. The function $a(t)$, known as the expansion parameter, is the ratio of the proper distance between any two galaxies at the initial time $t=0$ and the time $t$.

\begin{widetext}
During an infinitesimal duration of proper time $dt$ a photon travels the distance $d\eta=dt/a(t)$. It is convenient to use $\eta$, known as the conformal time coordinate, as the time parameter. Transforming to conformal time has the effect of shunting the time dependence into a conformal factor.  We furthermore consider a universe dominated by matter, in which case $a(\eta)=a_{\mathrm{max}}(1-\cos\eta)$ and the metric becomes
\begin{align}
ds^2=\frac{a_{\mathrm{max}}^2}{4}(1-\cos\eta)^2\!\left[-d\eta^2+d\chi^2
+\sin^2\!\chi\left(d\theta^2+\sin^2\!\theta\,d\phi^2\right)\right]\;,
\label{FLRW}
\end{align}
where $\eta$ runs between $0$ at the beginning of expansion to $2\pi$ at the end of recontraction. We wish to estimate the parameter~$a_{\mathrm{max}}$ which controls the maximum size the universe reaches before contraction commences.  Since $d\g_{\mu\nu}/da_\mathrm{max}=(2/a_\mathrm{max})\g_{\mu\nu}$, the operator~(\ref{Ptheta}) becomes
\begin{align}
\begin{split}
\hat{P}(a_{\mathrm{max}})
&=\frac{a_{\mathrm{max}}^3}{16}
\int_M d\eta\,d\chi\,d\theta\,d\phi\,(1-\cos\eta)^4\sin^2\!\chi\sin\theta
\,\g_{\mu\nu}\hT^{\mu\nu}\\
&=\frac{a_{\mathrm{max}}^5}{64}
\int_M d\eta\,d\chi\,d\theta\,d\phi\,(1-\cos\eta)^6\sin^2\!\chi\sin\theta
\Big[-\hat{T}^{\eta\eta}+\hat{T}^{\chi\chi}
+\sin^2\!\chi\big(\hat{T}^{\theta\theta}+\hat{T}^{\phi\phi}\sin^2\!\theta\big)\Big]\;.
\end{split}
\end{align}

It is interesting to note that since we are estimating a scale factor, the above integrand is proportional to the trace of the stress-energy tensor.  Thus, for a any field with a traceless stress-energy tensor, such as the free electromagnetic field, $\langle(\Delta\hP)^2\rangle$ vanishes, and we get no information about the scale factor.  This is an expression of the well-known scale invariance of the electromagnetic field.

To give another cosmological example, suppose we are interested in measuring
the cosmological constant $\Lambda$ in a de Sitter universe.  Again using the
conformal time coordinate, the line element can be written as
\begin{equation}
ds^2=\frac{3}{\Lambda}\sec^2\!\eta\,\Big(\mathord{-}d\eta^2+d\chi^2+\sin^2\!\chi\,(d\theta^2+\sin^2\!\theta\, d\phi^2)\Big)\;,
\label{deS}
\end{equation}
which gives us
\begin{equation}
\hP(\Lambda)
=-\frac{9}{2\Lambda^3}\int_M d\eta\,d\chi\,d\theta\,d\phi\,\sec^4\!\eta\sin^2\!\chi\sin\theta\,\g_{\mu\nu}\hT^{\mu\nu}\;.
\end{equation}
Again, because we are estimating a scale factor, the integrand is proportional to the trace of the stress-energy tensor, which vanishes for the free electromagnetic field.
\end{widetext}

These and other cosmological parameters represent overall scale factors of the universe.  As such, the conformally invariant electromagnetic field alone is not an adequate probe.  For example, it is only possible to measure the cosmological redshift of light if the atomic emission spectrum of its source is also known.  Indeed, Penrose has essentially argued that should all matter in the universe decay into photons, the scale of the universe would become unobservable and thus physically irrelevant~\cite{penrose}.  So a more useful calculation should include fermionic fields, which break this scale invariance.  Alternatively, a massive scalar or boson field would also yield a finite Cram\'er-Rao bound, since its stress-energy has nonvanishing trace.

A more down-to-earth application is that of a gravimeter.  One way to deal with this case is to approximate the near-earth spacetime by a Schwarzschild metric, and assume the floor of the laboratory has constant Schwarzschild coordinate radius; then the problem of gravimetry becomes one of estimating the Schwarzschild mass of the Earth.  This problem certainly lends itself to our method, provided the appropriate stress-energy correlations in a Schwarzshild
background are calculated.  Of course, the same calculation would also be applicable to a black hole.

Our method is also applicable to estimation of dynamical quantities of a probe in flat spacetime, such as acceleration, rotation, etc.  One approach is to assume such dynamics are due to coupling with a nongravitational, classical
field.  The locally covariant approach then ``applies, {\it mutatis mutandis}, also to this case"~\cite{bfv}, and so also does our Cram\'er-Rao bound.

\section{Conclusion}
\label{sec:conclusion}

In this paper we presented the quantum Cram{\'e}r-Rao lower bound for
the uncertainty in estimating parameters describing a spacetime
metric.  Our specific derivation applies for any quantum state on an
arbitrary globally hyperbolic manifold.  To demonstrate the utility of
our formalism, we applied it to estimation of metric components
and found uncertainty principles akin to those found by
Unruh~\cite{UNRUH:1986p358} using heuristic arguments.  We also
considered quantum estimation of gravitational-wave amplitude and
obtained generalizations of known quantum limits for laser
interferometers such as LIGO.

\appendix
\section{Category theory framework}
\label{sec:categorytheory}

In this appendix we briefly review category theory as employed by Brunetti, Fredenhagen, and Verch in \cite{bfv}.  A primary motivation is that it provides
a convenient way to rigorously define the change in a quantum
observable due to a spacetime perturbation, which we outline here.
To begin with, a {\it category\/} (or more precisely, a {\it concrete category}) consists of objects and functions between objects called {\it morphisms}.
A category of particular relevance for our purposes, given in~\cite{bfv}, is the following:
\begin{definition} $\mathfrak{Man}$ is the category whose objects are globally hyperbolic spacetimes and whose morphisms are isometric embeddings (or in other words inclusion maps).
\end{definition}
\noindent
We next consider quantum fields in those spacetimes, formulated in terms of $C^*$-algebras.  The relevant category of $C^*$-algebras is given in~\cite{bfv} as follows:
\begin{definition}
$\mathfrak{Alg}$ is the category whose objects are $C^*$-algebras and whose morphisms are injective *-homomorphisms.
\end{definition}

\begin{figure*}
%$$
%\begindc{\commdiag}[30]
%\obj(10,52)[man]{$\mathfrak{Man}$}
%\obj(70,52)[alg]{$\mathfrak{Alg}$}
%\obj(10,30)[mgc]{$(M,\mathring{\g})$}
%\obj(30,50)[np]{$(N_+,\g_{N_+})$}
%\obj(50,30)[mg]{$(M,\g)$}
%\obj(30,10)[nm]{$(N_-,\g_{N_-})$}
%\obj(70,30)[amgc]{$\scrA(M,\mathring{\g})$}
%\obj(90,50)[anp]{$\scrA(N_+,\g_{N_+})$}
%\obj(110,30)[amg]{$\scrA(M,\g)$}
%\obj(90,10)[anm]{$\scrA(N_-,\g_{N_-})$}
%\mor{np}{mgc}{$\psi^+_{\circ}$}[\atright,\solidarrow]
%\mor{np}{mg}{$\psi^+_{\g}$}
%\mor{nm}{mgc}{$\psi^-_{\circ}$}
%\mor{nm}{mg}{$\psi^-_{\g}$}[\atright,\solidarrow]
%\mor{mg}{amgc}{$\scrA$}[\atright,\solidarrow]
%\mor{anp}{amgc}{$\alpha_{\psi^+_{\circ}}$}[\atright,\solidarrow]
%\mor{anp}{amg}{$\alpha_{\psi^+_{\g}}$}
%\mor{anm}{amgc}{$\alpha_{\psi^-_{\circ}}$}
%\mor{anm}{amg}{$\alpha_{\psi^-_{\g}}$}[\atright,\solidarrow]
%\mor(4,4)(56,4)[0,0]{}[\atright,\solidline]
%\mor(56,4)(56,56)[0,0]{}[\atright,\solidline]
%\mor(56,56)(4,56)[0,0]{}[\atright,\solidline]
%\mor(4,56)(4,4)[0,0]{}[\atright,\solidline]
%\mor(62,56)(62,4)[0,0]{}[\atright,\solidline]
%\mor(62,4)(118,4)[0,0]{}[\atright,\solidline]
%\mor(118,4)(118,56)[0,0]{}[\atright,\solidline]
%\mor(118,56)(62,56)[0,0]{}[\atright,\solidline]
%\cmor((78,28)(90,18)(102,28)(102,30)(102,32)(90,42)(78,32)) \pdown (99,30){$\beta_\g$}[\dashArrow]
%%\mor(79,33)(78,32)[0,0]{}[\atright,\solidarrow]
%\enddc\;
%$$
\includegraphics{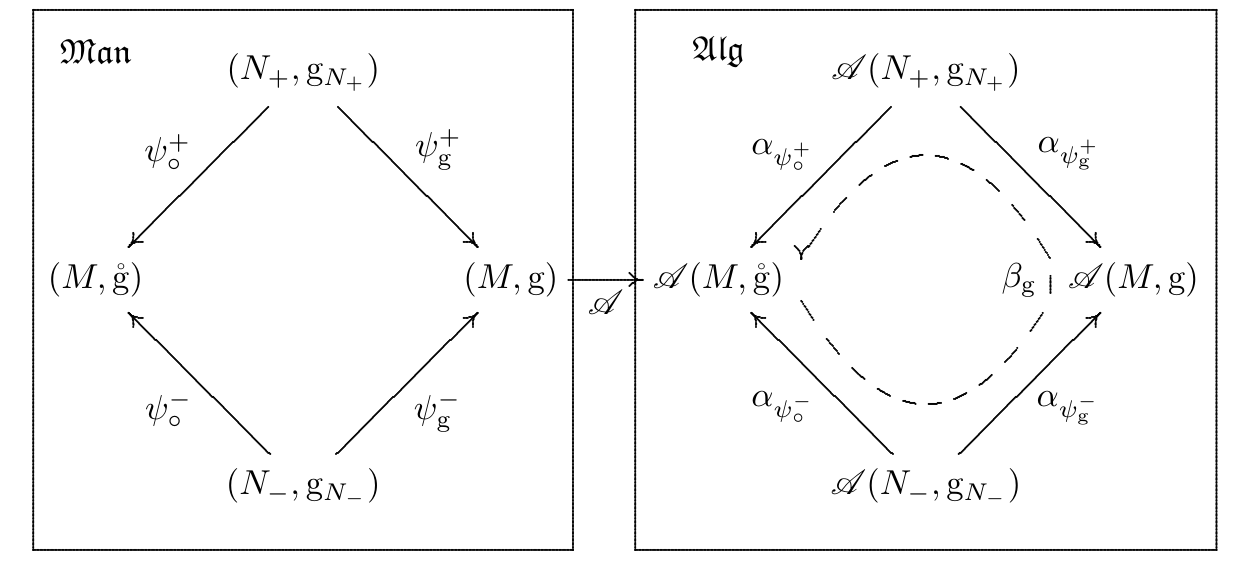}
\caption{Diagram of the perturbed spacetime, along with the embedded subregions $N_\pm$, the locally covariant quantum field theory thereon, and the relevant morphisms, where $\mathring{\g}=\g^{(0)}$ and $\g=\g^{(s)}$.}
   \label{fig:lcqft}
\end{figure*}

To associate $C^*$-algebras to our spacetimes, we use a covariant functor from the spacetimes to $C^*$-algebras.
A {\it functor\/} is a function between categories which maps objects to objects and morphisms to morphisms, such that the identity maps to the identity and compositions map to compositions.  A {\it covariant functor\/} is a functor that maps domains to domains and images to images (pictorially, it preserves the directions of morphism arrows).  Thus we arrive at the following~\cite{bfv}:
\begin{definition}
A locally covariant quantum field theory is a covariant functor,
\begin{equation*}
\scrA:\mathfrak{Man}\rightarrow\mathfrak{Alg}\;.
\end{equation*}
\end{definition}
\noindent
Note that a locally covariant quantum field theory is local in the sense that it maps submanifolds of manifolds to subalgebras of the corresponding algebras.

We further require that any causal, locally covariant quantum field theory obey the following axiom~\cite{bfv}:
\begin{axiom}\emph{(Time-Slice Axiom)}
If $\scrA$ is a locally covariant quantum field theory, $(N,g),(M,g)\in\mathfrak{Man}$, and $\psi\in\hom\big((N,g),(M,g)\big)$ such that $\psi(N,g)$ contains a Cauchy surface of $(M,g)$, then $\scrA(\psi)(\scrA(N,g))=\scrA(M,g)$.
\end{axiom}
\noindent
In other words, the algebra associated with a Cauchy surface of a manifold determines the algebra associated with the entire manifold.

The perturbed spacetime discussed above, along with the embedded subregions $N_\pm$, the locally covariant quantum field theory thereon, and the relevant morphisms, can all be represented by the diagram in Fig.~\ref{fig:lcqft}, where $\mathring{\g}=\g^{(0)}$ and $\g=\g^{(s)}$.  Note that the assumption of the time-slice axiom implies the morphisms $\scrA(N_\pm,\g_{N_\pm})\rightarrow\scrA(M,g)$ are bijective.  Therefore, the arrows in the right-hand side of the diagram are invertible.  This allows these morphisms to be composed in such a way as to construct an automorphism on $\scrA(M,\g^{(0)})$:
\begin{align}
\beta_{\g} = \alpha_{\psi_{\circ}^-}\circ\alpha^{-1}_{\psi_{\g}^-}\circ\alpha_{\psi_{\g}^+}\circ\alpha^{-1}_{\psi_{\circ}^+}\;.
\end{align}

By the Gelfand-Neimark-Segal construction~\cite{gelfand,segal,haag:1996}, every $C^*$-algebra admits a linear *-representation $\pi$ by bounded operators on a Hilbert space.  Thus $\beta_\g$ induces an automorphism on Hilbert-space operators.  Thus any operator $\hat{A}=\pi(A)$, where $A\in\scrA(M,\g^{(0)})$, i.e., any operator associated with our fiducial spacetime, can be said to evolve under our $s$-parametrized spacetime perturbation into $\hat{A}(s)=\pi(\beta_{\g^{(s)}}A)$.  This process is termed \emph{relative Cauchy evolution}~\cite{bfv}.

\section{Coordinate independence}
\label{sec:coordindep}

In this appendix we derive Eq.~(\ref{diffeo}), and thus the coordinate independence of $(\Delta\hP)^2$, without
the assumption of a single coordinate patch covering $K$.  This assumption is not valid, for example, when the interior of $K$ is not homeomorphic to $\mathbb{R}^4$.  More generally, it is often simply more convenient to use multiple coordinate patches.

For the purposes of this proof, however, we avoid explicitly juggling multiple coordinate transition maps by
considering each to be locally induced by a global diffeomorphism $\varphi^{(s)}:M\rightarrow M$ that depends continuously on the parameter $s\in[0,1]$.
This induces a pushforward $\varphi^{(s)}_*$ of a contravariant tensor field
or (in the same direction) a pullback of the inverse diffeomorphism $((\varphi^{(s)})^{-1})^*$ of a covariant field, which we will also denote by $\varphi^{(s)}_*$.
Note if restricted to a coordinate patch $\varphi^{(s)}_*$ is related to the transformation $L^\mu_{\alpha'}(s)$ of Sec.~\ref{sec:compact} by $(\varphi^{(s)}_*v)_{\alpha'}=L^\mu_{\alpha'}(s)v_\mu$.  (And notice that in the classic index analysis of Sec.~\ref{sec:compact}, the prime serves double duty, denoting both this $s$-dependent coordinate transformation, and its $s=0$ instance.)

To further obviate the need for explicit coordinate charts, we use here Penrose's abstract index notation \cite{penrose:1968}, denoted by Latin indices.  Like coordinate component indices, the number of such subscripted/superscripted indices indicate tensor ranks, and repeated indices indicate tensor contractions.  But Penrose's abstract indices do not index coordinate components, rather they signify entire tensors.  For example, $\g_{\mu\nu}\in\mathbb{R}$ while $\g_{ab}\in T^*M\otimes T^*M$.  Thus Penrose's abstract indices provide all the convenience of indices without any of the commitment to coordinates.  In these terms, our objective is to show
\begin{align}
\begin{split}
\int_K&\left.d\mu_{\varphi^{(s)}_*\g^{(s)}}\varphi^{(s)}_*T^{ab}\frac{d}{ds}\varphi^{(s)}_*\g_{ab}^{(s)}\right|_{s=0}\\
&=\int_K \left.d\mu_{\g^{(s)}}T^{ab}\frac{d}{ds}\g_{ab}^{(s)}\right|_{s=0}\;.
\end{split}
\end{align}

Assuming that $\nabla_a\hT^{ab}=0$, we have
\begin{widetext}
\begin{align}
\begin{split}
\int_K d\mathring\mu'\,\hat{T}^{\prime ab}\left.\frac{d}{ds}\right|_0\varphi_*^{(s)}\g^{(s)}_{ab}
&=\int_K d\mathring\mu'\,\hat{T}^{\prime ab}\left.\frac{d}{ds}\right|_0\phi_*^{(s)}\g^{(s)\prime}_{ab}\\
&=\int_K d\mathring\mu'\,\hat{T}^{\prime ab}\left[\left(\left.\frac{d}{ds}\right|_0\g^{(s)}_{ab}\right)'
+\left.\frac{d}{ds}\right|_0\phi_*^{(s)}\g^{(0)\prime}_{ab}\right]\\
&=\int_K d\mathring\mu'\,\hat{T}^{\prime ab}\left[\left(\left.\frac{d}{ds}\right|_0\g^{(s)}_{ab}\right)'+\nabla'_a X'_b+\nabla'_b X'_a\right]\\
&=\int_K d\mathring\mu\,\hat{T}^{ab}\left[\left.\frac{d}{ds}\right|_0\g^{(s)}_{ab}+\nabla_a X_b+\nabla_b X_a\right]\\
&=\int_K d\mathring\mu\,\left[\hat{T}^{ab}\left.\frac{d}{ds}\right|_0\g^{(s)}_{ab}
+2(\nabla_a(\hat{T}^{ab}X_b) -(\nabla_a\hat{T}^{ab}X_b))\right]\\
&=\int_K d\mathring\mu\,\hat{T}^{ab}\left.\frac{d}{ds}\right|_0\g^{(s)}_{ab}
+2\int_K d\mathring\mu\,\nabla_a(\hat{T}^{ab}X_b)\\
&=\int_K d\mathring\mu\,\hat{T}^{ab}\left.\frac{d}{ds}\right|_0\g^{(s)}_{ab}
+2\int_{\partial K} d\lambda\,n_a\hT^{ab}X_b\;,
\end{split}
\end{align}
\end{widetext}
where
$\phi^{(s)}=\varphi^{(s)}\circ(\varphi^{(0)})^{-1}$, $X^\prime_a$ generates $\phi^{(s)}$, $d\lambda$ is the surface element induced on the boundary of $K$, $n_a$ is the corresponding surface normal,
and primes denote $\phi^{(0)}_*$.
Then neglecting the boundary term,
as explained in Sec.~\ref{sec:compact},
we achieve the desired diffeomorphism-independence.

Note that $\phi^{(s)}$ above corresponds to $\phi^{(s)}$ in the proof of Theorem 4.2 in \cite{bfv}.
That theorem implies that $\nabla_a\hT^{ab}$ must vanish if the above integral is invariant with respect to the transformation
$\phi^{(s)}_*$ of the metric.  The above result is a generalization of the converse.

\section{Commutators and vacuum correlation functions for the electromagnetic field}
\label{sec:commcorr}

The field operators for the free electric and magnetic fields can be written as (we use cgs Gaussian units with $c=1$)
\begin{align}
\hvecE(\vect{x},t)
&=\hvecE^{(+)}(\vect{x},t)+\hvecE^{(-)}(\vect{x},t)\;,\\
\hvecB(\vect{x},t)
&=\hvecB^{(+)}(\vect{x},t)+\hvecB^{(-)}(\vect{x},t)\;,
\end{align}
where the positive- and negative-frequency parts of the fields are given by
\begin{widetext}
\begin{align}
\hvecE^{(+)}=\hvecE^{(-)\dagger}
&=i\sum_\sigma\int\dthreek\,\sqrt{2\pi\hbar\omega}\,\ha_\ks\vect{e}_\ks
e^{i\omega(\vect{n}\cdot\vect{x}-t)}\;,\\
\hvecB^{(+)}=\hvecB^{(-)\dagger}
&=i\sum_\sigma\int\dthreek\,\sqrt{2\pi\hbar\omega}\,\ha_\ks\vect{n}\times\vect{e}_\ks
e^{i\omega(\vect{n}\cdot\vect{x}-t)}\;.
\end{align}
Here $\vect{k}=\omega\vect{n}$ is the wave vector ($\omega=|\vect{k}|$ is the angular frequency and $\vect{n}$ a unit vector), and $\ha_\ks$ and $\vect{e}_\ks$ are the annihilation operator and unit (transverse) polarization vector for the plane-wave mode with wave vector $\vect{k}$ and polarization $\sigma$.  The creation and annihilation operators satisfy the canonical commutator,
\begin{align}
\big[\ha_\ks,\ha_\ksprime^\dagger\big]=(2\pi)^3\delta(\vect k-\vect k')\delta_{\sigma\sigma'}\;.
\end{align}
The Hamiltonian for the electromagnetic field is
\begin{align}\label{eq:EMH}
\hH
=\int d^3x\,\no{\hT^{00}}
=\frac{1}{8\pi}\int d^3x\,\no{\hvecE\cdot\hvecE+\hvecB\cdot\hvecB}
=\frac{1}{2\pi}\int d^3x\,\hvecE^{(-)}\cdot\hvecE^{(+)}
=\sum_\sigma\int\dthreek\,\hbar\omega a_{\ks}^\dagger a_{\ks}\;.
\end{align}
where the integral extends over all space.

The positive- and negative-frequency parts of the fields have the free-field commutators,
\begin{align}
\big[\hE_j^{(+)}(\vect x,t),\hE_k^{(-)}(\vect x',t')]
=\big[\hB_j^{(+)}(\vect x,t),\hB_k^{(-)}(\vect x',t')]
&=2\pi\hbar
\left(-\delta_{jk}\frac{\partial^2}{\partial t^2}+\frac{\partial^2}{\partial x_j\partial x_k}\right)
\int\dthreek\,\frac{1}{\omega}e^{i\omega[\vect{n}\cdot(\vect x-\vect x')-(t-t')]}\;,\\
\big[\hE_j^{(+)}(\vect x,t),\hB_k^{(-)}(\vect x',t')]
=-\big[\hB_j^{(+)}(\vect x,t),\hE_k^{(-)}(\vect x',t')]
&=2\pi\hbar\,\epsilon_{jkl}\frac{\partial^2}{\partial t\,\partial x_l}
\int\dthreek\,\frac{1}{\omega}e^{i\omega[\vect{n}\cdot(\vect x-\vect x')-(t-t')]}\;.
\end{align}
The integral on the right evaluates to
\begin{align}
\int\dthreek\,\frac{1}{\omega}e^{i\omega[\vect{n}\cdot(\vect x-\vect x')-(t-t')]}
=-\frac{i}{4\pi}G(\vect x-\vect x',t-t')+\frac{1}{2\pi^2}D(\vect x-\vect x',t-t')\;,
\end{align}
where
\begin{align}\label{eq:G}
G(\vect x,t)
=-4\pi\,\mbox{Im}\int\dthreek\,\frac{1}{\omega}e^{i\omega(\vect{n}\cdot\vect x-t)}
=\frac{\delta\big(t-|\vect x|\big)-\delta\big(t+|\vect x|\big)}{|\vect x|}
\end{align}
is the difference between retarded and advanced Green functions, i.e., the solution of the homogeneous wave equation for an incoming spherical wave that reflects off the origin and becomes an outgoing spherical wave,
and
\begin{align}\label{eq:D}
D(\vect x,t)
=2\pi^2\,\mbox{Re}\int\dthreek\,\frac{1}{\omega}e^{i\omega(\vect{n}\cdot\vect x-t)}
=\text{p.v.}\frac{1}{-t^2+|\vect x|^2}=\text{p.v.}\frac{1}{(\Delta s)^2}
\end{align}
is the principal value of the inverse of the invariant interval.  From these follow the free-field commutators and vacuum correlators~\cite{greiner}:
\begin{align}
\begin{split}\label{eq:EorBcomm}
\big[\hE_j(\vect x,t),\hE_k(\vect x',t')]&=\big[\hB_j(\vect x,t),\hB_k(\vect x',t')]\\[3pt]
&=2i\,\mbox{Im}\big(\big[\hE_j^{(+)}(\vect x,t),\hE_k^{(-)}(\vect x',t')]\big)\\
&=i\hbar
\left(\delta_{jk}\frac{\partial^2}{\partial t^2}-\frac{\partial^2}{\partial x_j\partial x_k}\right)
G(\vect x-\vect x',t-t')\;,
\end{split}\\
\begin{split}\label{eq:EwithBcomm}
\big[\hE_j(\vect x,t),\hB_k(\vect x',t')]&=-\big[\hB_j(\vect x,t),\hE_k(\vect x',t')]\\[3pt]
&=2i\,\mbox{Im}\big(\big[\hE_j^{(+)}(\vect x,t),\hB_k^{(-)}(\vect x',t')]\big)\\
&=-i\hbar\,\epsilon_{jkl}\frac{\partial^2}{\partial t\partial x_l}
G(\vect x-\vect x',t-t')\;,
\end{split}\\
\begin{split}\label{eq:EorBcorr}
\frac12\big\langle0\big|[\hE_j(\vect x,t)\hE_k(\vect x',t')+\hE_k(\vect x',t')\hE_j(\vect x,t)]\big|0\big\rangle
&=\frac12\big\langle0\big|[\hB_j(\vect x,t)\hB_k(\vect x',t')+\hB_k(\vect x',t')\hB_j(\vect x,t)]\big|0\big\rangle\\[3pt]
&=\mbox{Re}\big(\big[\hE_j^{(+)}(\vect x,t),\hE_k^{(-)}(\vect x',t')]\big)\\
&=\frac{\hbar}{\pi}\left(-\delta_{jk}\frac{\partial^2}{\partial t^2}+\frac{\partial^2}{\partial x_j\partial x_k}\right)
D(\vect x-\vect x',t-t')\;,
\end{split}\\
\begin{split}\label{eq:EwithBcorr}
\frac12\big\langle0\big|[\hE_j(\vect x,t)\hB_k(\vect x',t')+\hB_k(\vect x',t')\hE_j(\vect x,t)]\big|0\big\rangle
&=-\frac12\big\langle0\big|[\hB_j(\vect x,t)\hE_k(\vect x',t')+\hE_k(\vect x',t')\hB_j(\vect x,t)]\big|0\big\rangle\\[3pt]
&=\mbox{Re}\big(\big[\hE_j^{(+)}(\vect x,t),\hB_k^{(-)}(\vect x',t')]\big)\\
&=\frac{\hbar}{\pi}\,\epsilon_{jkl}\frac{\partial^2}{\partial t\,\partial x_l}
D(\vect x-\vect x',t-t')\;.
\end{split}
\end{align}

The field quadratures~(\ref{eq:hX1}) and~(\ref{eq:hX2}) for a nearly plane-wave mean field can be evaluated in the Fourier domain as follows ($a=1,2$):
\begin{align}
\begin{split}
\hX_a=\frac{1}{4\pi}\int d^4x\,E_a(\vect x,t)\hE_y(\vect x,t)
&=\frac{1}{4\pi}\int dt\,d^3x\,E_a^{(-)}(\vect x,t)\hE_y^{(+)}(\vect x,t)+\mbox{H.c.}\\
&=\frac12\tau\int\dthreek\,\hbar\omega\sum_{\sigma,\sigma'}\alpha_{a;\ks}\ha_{\vect k\sigma'}
(\vect e_y\cdot\vect e^*_\ks)(\vect e_y\cdot\vect e_{\vect k\sigma'})+\mbox{H.c.}\\
&=\frac12\tau\int\dthreek\,\hbar\omega\,\alpha_{a,\vect k}\ha_{\vect ky}+\mbox{H.c.}
\;.
\end{split}
\end{align}
\end{widetext}
In the first line we discard counter-rotating terms that average to nearly zero over the temporal integral; in the second line, we insert the Fourier transforms of the field operators and the wave forms and do the temporal integral over the duration $\tau$ for which the mean fields are turned on; in the third line, we use the fact that $\alpha_{a;\ks}$ has support only for $\vect k$ pointing nearly in the $x$ direction, with polarization nearly along the $y$ direction, to restrict the two sums over polarization to $y$ linear polarization.

Using Eq.~(\ref{eq:alpha21}), we can write
\begin{align}\label{eq:XA}
\begin{split}
\hX_1&=\sqrt{\frac{\hbar C}{2}}(\hb+\hb^\dagger)\;,\\
\hX_2&=\sqrt{\frac{\hbar C}{2}}(-i\hb+i\hb^\dagger)\;,
\end{split}
\end{align}
i.e., $\hb=(\hX_1+i\hX_2)/\sqrt{2\hbar C}$, where
\begin{align}
\hb=\frac{\tau}{\sqrt{2\hbar C}}\int\dthreek\,\hbar\omega\,\alpha^*_{1,\vect k}\ha_{\vect ky}
\end{align}
[the constant $C$ is given in Eq.~(\ref{eq:C})].  One can verify that the quadrature components obey the commutation relation~(\ref{eq:QUADcomm}) or, equivalently, that $\hb$ and $\hb^\dagger$ satisfy the canonical bosonic commutation relation, $[\hb,\hb^\dagger]=1$.

\acknowledgments
TGD thanks K.~Fredenhagen and support from DESY, as well as R.~Verch and support from the University of Leipzig.  We also thank A.~Kempf, T.~C. Ralph, J.~Doukas, C.~Fewster, L.~H. Ford, and S.~Carlip for useful discussions.  This work was supported in part by US National Science Foundation Grant Nos.~PHY-1314763 and PHY-1521016.  This work includes contributions of the National Institute of Standards and Technology, which are not subject to U.S. copyright.

\end{document}